\documentclass[aps,prb,reprint,twocolumn,longbibliography,floatfix]{revtex4-1}
\usepackage{graphicx}
\usepackage{amsmath,amssymb,amsfonts,bm,dsfont,units}
\usepackage[colorlinks,citecolor=blue,urlcolor=blue]{hyperref}

\newcommand{\vect}[1]{\boldsymbol{#1}}

\begin{document}

\title{Phase diagram of compressible and paired states in the quarter-filled Landau level}
\author{Misha Yutushui}
\affiliation{Department of Condensed Matter Physics, Weizmann Institute of Science, Rehovot 76100, Israel} 
\author{David F. Mross}
\affiliation{Department of Condensed Matter Physics, Weizmann Institute of Science, Rehovot 76100, Israel} 
\begin{abstract}
Quantum Hall plateaus at quarter fillings occur in GaAs wide quantum wells, hole-doped GaAs, and bilayer graphene. However, the interactions favoring incompressible states over compressible composite-Fermi liquids at such fillings are not well understood. We devise a method for the computation of the trial energies for Haldane pseudopotentials via Monte Carlo sampling. Applying it to the quarter-filled lowest Landau level, we find that tuning the third and fifth pseudopotential can stabilize anti-Pfaffian, Moore-Read, and f-wave states. The smallest deviations from pure Coulomb interactions are required by anti-Pfaffian, whose presence is indicated by daughter states in recent experiments of bilayer graphene at $\nu=\frac{3}{4}$.
\end{abstract}

\date{\today}
\maketitle

\section{Introduction and main result}
The fractional quantum Hall effect~\cite{Tsui_fqh_1982,Laughlin_fqh_1983,Haldane_fqh_1983,Halperin_fqh_1983} at even-denominators has been attracting widespread attention due to the possibility of hosting non-Abelian anyons.~\cite{Wen_Non_Abelian_1991,nayak_non-abelian_2008,Moore_nonabelions_1991,Read_paired_2000}
The best-known candidate for such a state is the half-filled plateau in the first excited Landau level of GaAs, i.e., at the filling factor $\nu=\frac{5}{2}$.~\cite{Willett_observation_1987} The gap at this filling factor is attributed to the pairing of composite fermions~\cite{Read_paired_2000} -- electrons bound to two flux quanta, which exactly cancel the electron charge at half-filling.~\cite{Jain_composite-fermion_1989,Jain_composite_2007} Numerical studies of the half-filled Landau level found that Coulomb interactions favor a compressible composite-Fermi liquid (CFL) in the lowest Landau level (0LL). In the first excited Landau level (1LL), a paired state is stabilized instead.~\cite{Morf_transition_1998,Rezayi_incompressible_2000} The transition from metallic to paired state can be induced by reducing the first Haldane pseudopotential,~\cite{Haldane_fqh_1983} which suppresses the repulsion between electrons at short distances.

Further numerical studies~\cite{Wojs_landau_level_2010,Storni_fractional_2010,feiguin_density_2008,Feiguin_spin_2009,Peterson_Finite_Layer_Thickness_2008,Rezayi_breaking_2011,Pakrouski_phase_2015} confirmed these results and focused on distinguishing the leading candidates: the Moore-Read~\cite{Moore_nonabelions_1991} state and its particle-hole conjugate, known as anti-Pfaffian.~\cite{Levin_particle_hole_2007,Lee_particle_hole_2007} However, neither of these states is compatible with thermal transport measurements in GaAs quantum wells at $\nu=\frac{5}{2}$.~\cite{Banerjee_observation_2018,Dutta_Distinguishing_2022,Dutta_Isolated_2022,Melcer_Absent_2022,Paul_Thermal_2023,Melcer_Heat_2024,Paul_Thermal_2024,Dutta_novel_2022} These experiments instead indicate a particle-hole symmetric topological order.~\cite{Son_is_2015} Such a phase has never been found numerically, and disorder may be essential for understanding its origin.~\cite{Mross_theory_2018,Wang_topological_2018,Lian_theory_2018} Still, the finding that 1LL is conducive to pairing at half filling appears robust and extends to other quantum Hall platforms, such as bilayer graphene.~\cite{Ki_bilyaer_graphene_2014,Kim_bilayer_graphene_2015,Li_bilayer_graphene_2017,Zibrov_Tunable_bilayer_graphene_2017,Assouline_Energy_Gap_bilayer_graphene_2024}

\begin{figure}
    \centering
    \includegraphics[width=\linewidth]{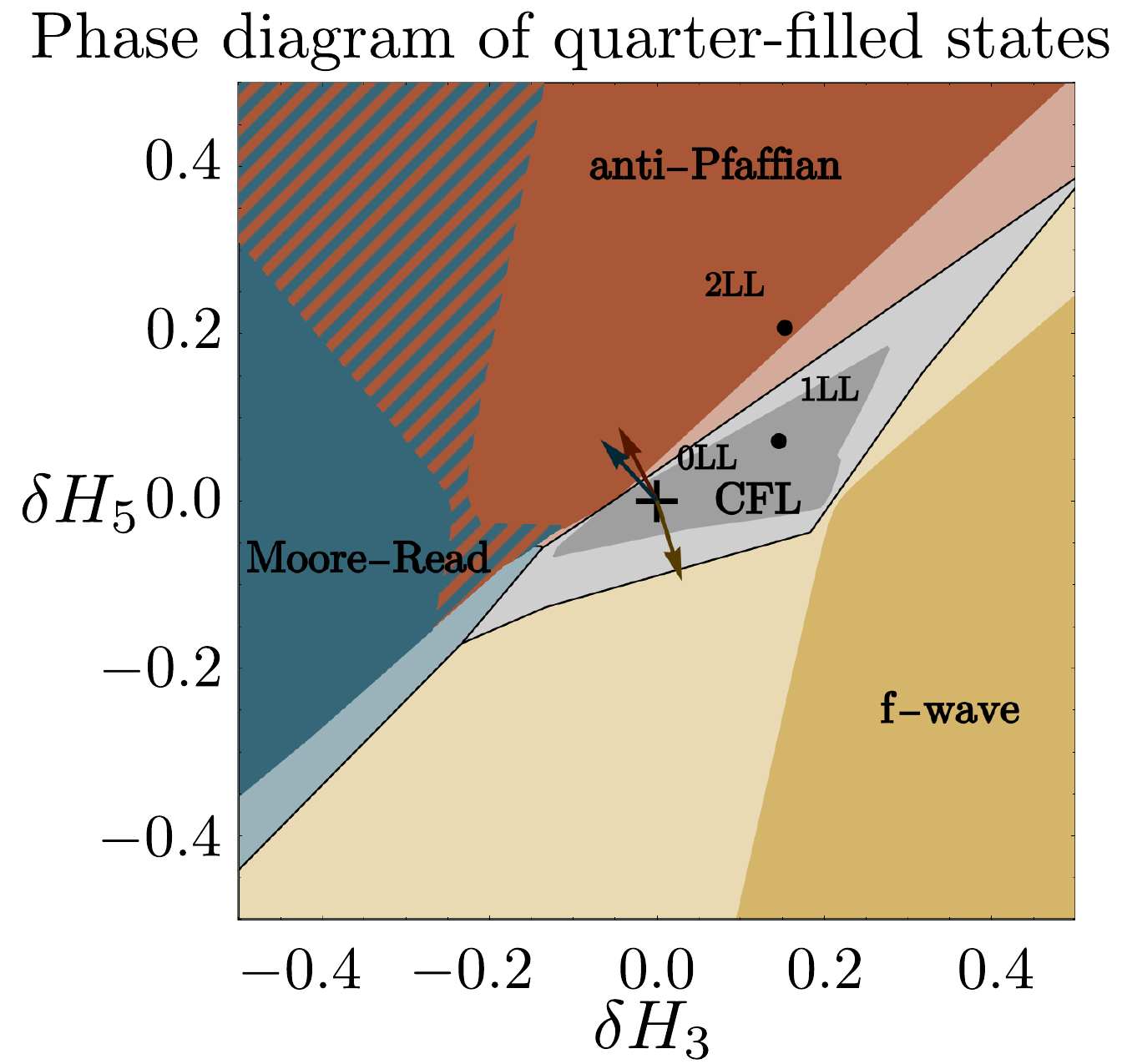}\\ 
    \caption{Phase diagram of the quarter-filled lowest Landau level with Coulomb interactions perturbed by $L=3,5$ Haldane pseudopotentials. Energies of trial wavefunctions for composite Fermi liquids and paired states were extrapolated to the thermodynamic limit to identify the favored phase. We colored regions in the $H_3,H_5$ plane according to the state with the lowest energy, with darker shades indicating a confidence level above $99\%$. The hatched region exhibits a close competition between trial states for Moore-Read and anti-Pfaffian pairing. The pure Coulomb point is indicated by $\boldsymbol{+}$. For reference, we also marked the $H_3,H_5$ values of the 1LL and 2LL. The arrows indicate the direction in which the paired states are favored, and the systems with $N_e=10$ electrons are studied via exact diagonalization in Fig.~\ref{fig.overlaps}.
    }
    \label{fig.phase_quart}
\end{figure}
Closely related paired states can occur in quarter-filled Landau levels.~\cite{Read_paired_2000} Here, binding four flux quanta to each electron yields charge-neutral composite fermions. Plateaus at quarter filling have been observed in GaAs-wide quantum wells~\cite{luhman_Observation_2008,shabani_evidence_2009} and, more recently, in hole-doped GaAs~\cite{Wang_even_3_4_2022,Wang_quarter_hole_2023} and bilayer graphene.~\cite{Kumar_Quarter_2024} Surprisingly, all of those states arise in the 0LL, although Landau-level mixing may be significant. The conditions that favor pairing at quarter filling are less well understood than those at half filling. Exact diagonalization studies are limited to $N_e \leq 12$ particles, and the fate of these systems in the thermodynamic limit remains unclear.~\cite{Papic_quarter_2009,Sharma_CF_pairing_quart_2024} Still, Monte Carlo studies of large systems indicate that an f-wave paired state\cite{Wen_Non_Abelian_1991} may be favored in wide quantum wells at quarter filling.~\cite{Papic_quarter_2009,Faugno_Prediction_2019,Sharma_CF_pairing_quart_2024} Fixed-phase diffusion Monte-Carlo approach finds that anti-Pfaffian can be stabilized in the presence of strong Landau level mixing.~\cite{Zhao_CF_pairing_LLM_2023}

\begin{figure}[t]
    \centering
    \includegraphics[width=1\linewidth]{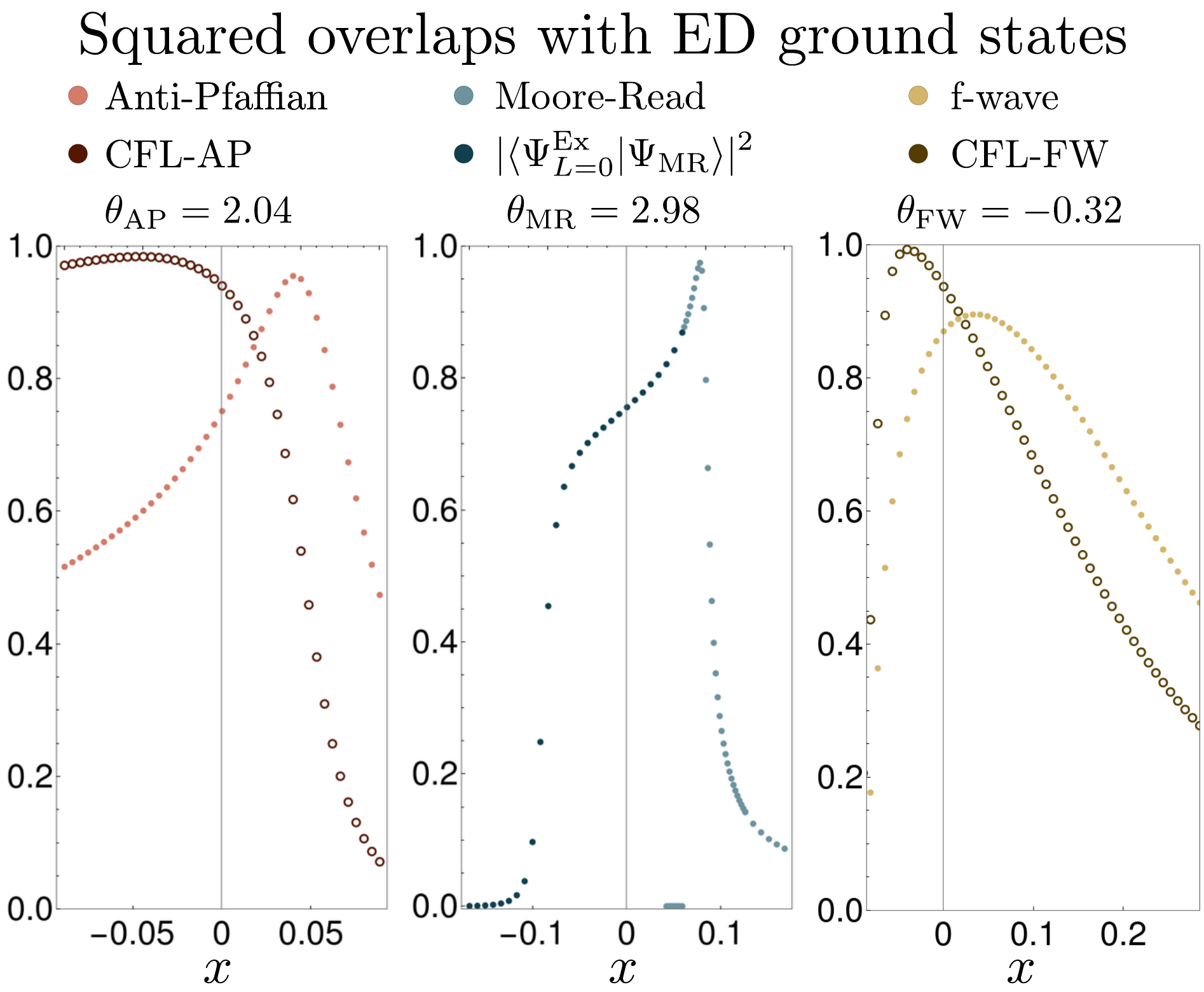}
    \caption{The overlaps of the ground state $|GS(x)\rangle$ of Eq.~\eqref{eqn:hx} with the trial states at $\nu=\frac{1}{4}$ for $N_e=10$. The angles $\theta_\Psi$ encode the directions in the $\delta H_3-\delta H_5$ plane indicated by arrows in Fig.~\ref{fig.phase_quart}, that favor the paired state $\Psi$. For the anti-Pfaffian and f-wave cases, modifying the third and fifth Haldane pseudopotentials along these directions results in transitions from compressible CFL state for pure 0LL Coulomb interactions $(x=0)$ to the paired state at positive $x$. At the Moore-Read shift, the ground state at $x=0$ has non-zero angular momentum. Since $N_e=10$ does not correspond to a filled shell~\cite{Rezayi_fermi-liquid-like_1994} of composite fermions at this shift, this finding is consistent with a composite Fermi liquid ground state. Here, we show the overlap of the trial state with the lowest zero angular momentum state $|GS_{L=0}(x)\rangle$, in addition to the one with the overall ground state. For all three shifts, the squared overlap with paired trial states peaks at positive $x$ and is above $90\%$.
    }
    \label{fig.overlaps}
\end{figure}

In this work, we systematically analyze the competition between compressible and paired states in a quarter-filled Landau level with general interactions. To achieve this, we develop an efficient method for evaluating the trial wavefunction energies for arbitrary choices of Haldane pseudopotentials directly with Monte Carlo simulations. Our numerical findings are summarized by the phase diagram in Fig.~\ref{fig.phase_quart}. It is obtained by computing the energy of different trial states at system sizes up to $N_e = 56$ particle and extrapolating to the thermodynamic limit.  We find that the \textit{third and fifth} Haldane pseudopotentials yield the strongest distinction between competing states at this filling. In the lowest Landau level with pure Coulomb interactions we find that the composite-Fermi liquid is favored. The paired anti-Pfaffian, Moore-Read, and f-wave states can all be stabilized by small changes in pseudopotentials; see Fig.~\ref{fig.overlaps}. In particular, the anti-Pfaffian paired state is close to the Coulomb point in the 0LL. This finding is consistent with recent experimental results~\cite{Kumar_Quarter_2024} of $\nu=\frac{3}{4}$ plateaus in bilayer graphene, where plateaus are accompanied by `daughter' states~\cite{Levin_collective_2009,Yutushui_daughters_2024,Zheltonozhskii_daughters_2024,Zhang_hierarchy_2024} that support the anti-Pfaffian pairing.

To verify that an energy difference between the specific trial states is indicative of the ground state, we performed complementary exact diagonalization studies of moderate system sizes. In Fig.~\ref{fig.overlaps} we compare the ground state of a Coulomb Hamiltonian with modified third and fifth pseudopotential to trial states of CFL and the three main pairing channels. In all three cases, we find that the overlap between the ground state and paired trial states peaks at a value close to unity in accordance with Fig.~\ref{fig.phase_quart}.

The rest of this work is organized as follows: In Sec.~\ref{sec.energies}, we briefly introduce the key players: (i) Haldane pseudopotentials, (ii) Real-space harmonics, and (iii) Static structure factors. Sec.~\ref{sec.map} describes the numerical methods developed in this work. After providing a general mapping between Haldane pseudopotential and real-space harmonics, we introduce the refinements necessary to obtain numerically stable results. In Sec.~\ref{subsec.laughlin}, we benchmark this approach by comparing trial energies from Monte Carlo to those obtained via exact diagonalization. In Sec.~\ref{subsec.cfl}, we present results for CFLs at different shifts, filling factors, and Landau levels. Sec.~\ref{sec.paired} contains our main results on paired states at half and quarter filling. We conclude in Sec.~\ref{sec.discussion} by discussing the implications and possible extensions of our work. Appendices contain additional details: In Appendix~\ref{app.higher}, we derive the real-space form of higher Landau-level pseudopotentials. Appendix~\ref{app.background} discusses finite size correction and introduces the normalization of energies throughout the paper. Appendix~\ref{app.projection} contains details of the Landau level projection. Appendix~\ref{app.overlaps} lists overlaps of paired states and CFLs.

\section{Energies in Exact Diagonalization and Monte Carlo}
\label{sec.energies}

Arbitrary two-body interactions between particles that occupy a single Landau level can be efficiently parametrized by a set of Haldane pseudopotentials~\cite{Haldane_fqh_1983} as
\begin{align}\label{eq.H_second}
    \mathcal{H}  = \sum_{L=0}^{2q} H_L \hat{\cal P}_{L},
\end{align}
where $\hat{\cal P}_{L}$ is a two-body operator projecting to relative angular momentum $L$.  On a sphere with a monopole of strength $q$, the number of states in the lowest Landau level is $2q+1$, and Haldane pseudopotentials with $L=0,\ldots,2q$ uniquely define any interaction. The known values of spherical pseudopotentials for Coulomb, $H^\text{$n$LL}_L(q)$, are derived in Appendix~\ref{app.higher}. (We will henceforth keep the dependence on $q$ implicit to lighten the notation.) Moreover, for a system of fermions (bosons), only $\hat{\cal P}_L$ for odd (even) $L$ have non-zero eigenvalues.

In numerical studies, one frequently chooses a specific interaction potential, encoded via $\{H^0_L\}$, as a reference point and varies a small number of pseudopotentials $\{\delta H_L\}$ to obtain a phase diagram like the one in Fig.~\ref{fig.phase_quart}. With exact diagonalization, one can compute the expectation values of any Hamiltonian $\mathcal{H}$ in a state $\Psi$ as $E_\Psi =\sum_{L} H_L \langle\hat{\cal P}_{L}\rangle_\Psi$~\cite{Wojs_3body_2005}; see also Ref.~\onlinecite{Jain_composite_2007}, in particular Appendix F for a review of the exact diagonalization method. However, this technique is limited to small systems, and it is often challenging to extrapolate to the thermodynamic limit. 

\subsection{Energies from Monte Carlo}

Given a trial wavefunction $\Psi(\vect r)$ and real-space potential $V(r)$, one can efficiently use Monte Carlo sampling (see Appending K of Ref.~\onlinecite{Jain_composite_2007} for details) to evaluate the integral 
\begin{align}\label{eq.energy_int_1}
    E_{V} = \frac{1}{2}\int |\Psi(\{\vect r_k\})|^2 \sum_{i\neq j}V(|\vec{r}_i - \vec{r}_j|)~.
\end{align}
This procedure works well for lowest-Landau level wave functions and moderately smooth potentials such as the Coulomb interaction $V(r)=\frac{e^2}{4\pi\epsilon |\vect r|}$.

The energy of higher Landau levels can be similarly computed if $\Psi$ is projected to the corresponding Landau level. However, most trial states and known methods of Landau-level projection are restricted to the lowest Landau level. The different single-particle orbitals in the $n$th Landau level compared to 0LL can alternatively be incorporated by a modified interaction. Approximate real-space potentials $V^\text{$n$LL}_\text{approx}(r)$ for this purpose have been constructed, e.g., in Refs.~\onlinecite{Park_possibility_1998,Toke_fqh_2005}, by requiring the lowest pseudopotentials to coincide. The price to pay is that $V^\text{$n$LL}_\text{approx}(r)$ is less smooth, and the Monte Carlo algorithm takes a longer time to converge. Thus, only the first few Landau levels are accessible.

\subsection{Static structure factors and energy}
\begin{figure}[t!]
     \centering
     \includegraphics[width=1\linewidth]{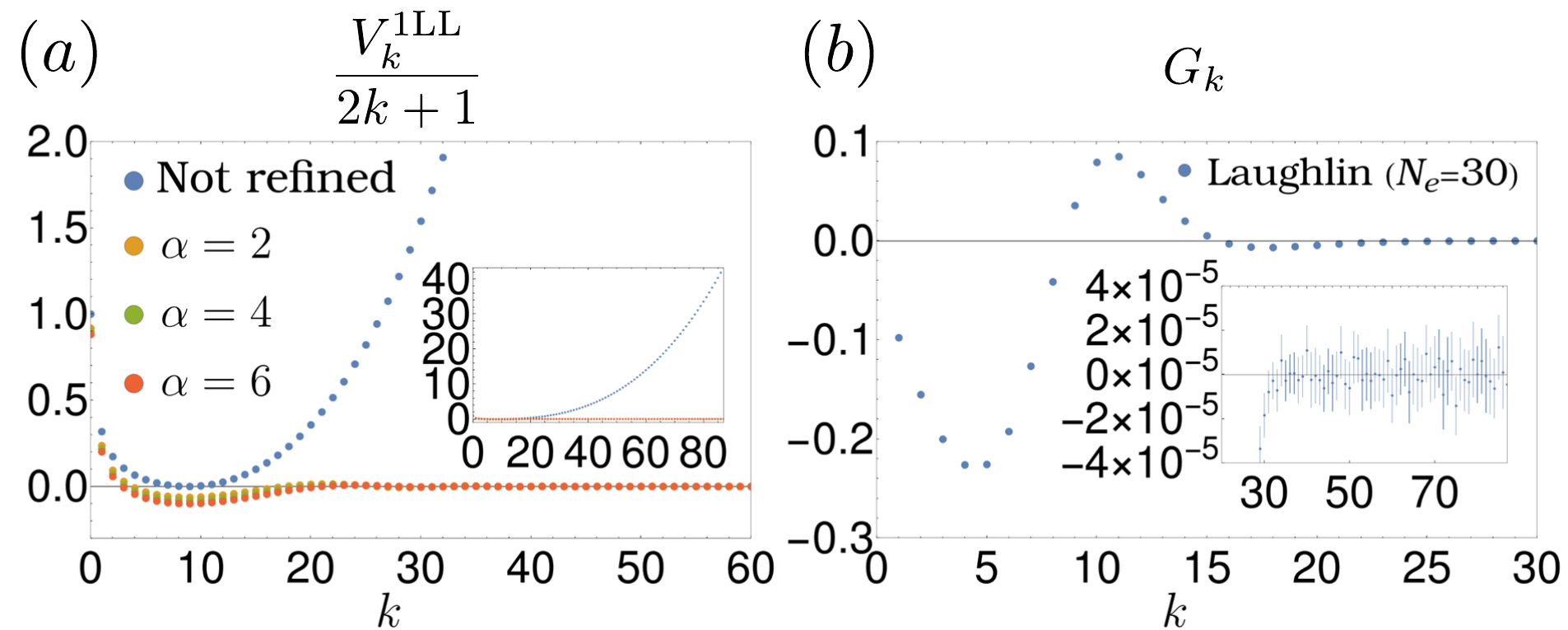}
     \caption{(a) The harmonics $V_k/(2k+1)$ encoding Coulomb interactions in the 1LL grow as $k^3$. The refined harmonics $\tilde V_k$ of Sec.~\ref{sec.refine} instead decay to zero. Different choices of the refinement parameter $\alpha$ in Eq.~\eqref{eq.opt} yield very similar coefficients. This behavior is shown here for $2l=87$ but is representative of general system sizes.
     (b) Expansion coefficients $G_k$ of the density-density correlation function obtained via Monte Carlo for a Laughlin state with $N_e=30$. The coefficients are close to zero for $k\gtrsim 15$ but exhibit noise due to finite sampling, as seen in the inset.  Without refinement, this noise is significantly amplified by the growing coefficient $V_k$ in Eq.~\eqref{eq.energy}. After refinement, the harmonics $\tilde{V}_k$ approach zero for large $k$, resulting in a numerically stable estimate for the energy.}
     \label{fig.vk_opt}
 \end{figure}

On a sphere of radius $R_s$, we define the distance between particles as the chordal distance $r=2R_s\sin(\theta/2)$, where $\theta \in [0,\pi]$ is the arc angle between their coordinates. For a spherically symmetric 0LL state $\Psi$, the energy Eq.~\eqref{eq.energy_int_1} can be expressed as 
\begin{align}\label{eq.E_integeral}
    E_V= \frac{N_e^2}{4} \int d\cos\theta\; G(\theta) V(\theta).
\end{align}
Here, $G(\theta)$ is the density-density correlation function
\begin{align}
    G(\theta) = \frac{2}{N^2_e}\int|\Psi(\{\vect r_i\})|^2 \sum_{i\neq j} \delta(
    \vec{r}_{i}\cdot\vec{r}_j-\cos\theta),
\end{align} 
normalized to approach unity for $\theta \gg N_e^{-1/2}$. Once this correlation function has been obtained, one can readily compute the energy for arbitrary potentials. We express the function $G(\theta)$ via a finite number of parameters by expanding it in terms of Legendre polynomials $P_k$, i.e., \begin{align}\label{eq.expnasion}
    G(\theta) = \sum_{k=0}^{2q} G_k P_{k}(\cos\theta).
\end{align}
(The moments with $k>2q$ vanish for a lowest-Landau-level wavefunction.) The expansion coefficients $G_k$ quickly tend to zero for smooth wavefunctions; see Fig.~\ref{fig.vk_opt}. We similarly expand the potential 
\begin{align}\label{eq.expnasion2}
    V(\theta) = \sum_{k=0}^{2q} V_k P_{k}(\cos\theta)~.
\end{align} Then, the integral Eq.~\eqref{eq.E_integeral} becomes
\begin{align}\label{eq.energy}
    E_V = \frac{N_e^2}{2}\sum_{k=0}^{2q}\frac{ V_kG_k }{2k+1}.
\end{align}
[To avoid confusion, we clarify that $V_k$ throughout this work denote the harmonics in the expansion Eq.~\eqref{eq.expnasion2}, while Haldane pseudopotentials are labeled by $H_L$.] We evaluate the $G_k$ directly from Monte Carlo via
\begin{align}
    G_k = \frac{2k+1}{N_e^2} \sum_{i\neq j}\left\langle P_k(\vec{r}_i\cdot \vec{r}_j)\right\rangle.
\end{align}
The zeroth coefficient encodes the total particle number $N_e=(1-G_0)^{-1}$. Apart from their use for computing trial energies, the expansion coefficients $G_k$ also directly determine the static structure factors
\begin{align}
    S_k = 1+\frac{N_e}{2k+1}(G_k-\delta_{k,0})~,
\end{align}
which can diagnose metallic states via singular behavior around $k =2k_F \approx \sqrt{N_e}$.~\cite{Kamilla_static_1997,Wojs_3body_2005} 

The energy in either of the equivalent forms in Eqs.~\eqref{eq.energy_int_1}, \eqref{eq.E_integeral}, or \eqref{eq.energy} can be evaluated efficiently with Monte Carlo for smooth potentials. The most prominent example is given by Coulomb interactions in 0LL, for which the harmonics are constant $V_k = R_s^{-1}$, which ensures rapid convergence of Eq.~\eqref{eq.energy}. The interactions $V^\text{$n$LL}_\text{approx}(r)$ are more oscillatory for $n>0$, and convergence is significantly slower. Below, we will show how interactions in higher Landau levels can be treated exactly and with substantially better numerical stability.

\section{Map between pseudopotentials and real-space harmonics.}
\label{sec.map}
The expression in Eq.~\eqref{eq.energy} allows computing the energy of any real-space potential expressed via the harmonics $V_k$. In particular, the energy of a trial state for any two-body Hamiltonian in the form of Eq.~\eqref{eq.H_second} can be evaluated by finding a set of $V_k$ representing $H_{L}$. For a set of Haldane pseudopotentials $H_L$ on a sphere with a monopole of strength $q$, the harmonics are
\begin{align}
   V_k = \sum_{L=0}^{2q}N_{k,L}(q) H_L~,
   \label{eqn.hltovk}
\end{align}
where the $2q+1$-dimensional transition matrix~\cite{wooten_haldane_2013} is 
\begin{align}
    N_{k,L}(q) =(-1)^L\frac{(2k+1)}{(2q+1)^2}(2(2q-L)+1)\nonumber\\\times
\left(
\begin{array}{ccc}
 q & k & q \\
 -q & 0 & q \\
\end{array}
\right)^{-2}
    \left\{
\begin{array}{ccc}
 q & q & k\\
 q & q & 2q-L  \\
\end{array}
\right\}~,
\label{eqn.nkl}
\end{align}
with $(\cdots)$  and $\{\cdots\}$ denoting 3$j$- and 6$j$-symbols, respectively. For Coulomb in 0LL, Eqs.~\eqref{eqn.hltovk} and \eqref{eqn.nkl} yield the previously quoted $k$-independent coefficients $V_k = R_s^{-1}$.

In Appendix~\ref{app.higher}, we show that, for arbitrary potentials, the coefficients $V_k$ in the $n$th Landau level can be obtained from those of the 0LL by
\begin{align}\label{eq.Vk$n$LL}
\begin{split}
    V^\text{$n$LL}_k&=V^\text{0LL}_k
\left(
\begin{array}{ccc}
 q & k & q \\
 n-q & 0 & q-n \\
\end{array}
\right)^{2}
    \left(
\begin{array}{ccc}
 q & k & q \\
 -q & 0 & q \\
\end{array}
\right)^{-2}\\
&\sim V^\text{0LL}_k k^{4n}\qquad\text{for $k\gg 1$}.
\end{split}~.
\end{align}
For the first Landau level, this expression reduces to
\begin{align}
    V^\text{1LL}_k=V^{0LL}_k \left(\frac{2q-k(k-1)}{2q}\right)^2~.
\end{align}
For $n>0$, in Eq.~\eqref{eq.energy}, the rapidly increasing $V^\text{$n$LL}_k$ multiplies any statistical error in the decaying coefficients $G_k$, which dramatically increases numerical noise; see Fig.~\ref{fig.vk_opt}. The resulting error becomes significant for large systems in the first and even moderate systems in higher Landau levels. Fortunately, the energy can be computed \textit{exactly} using refined coefficients $\tilde V_k$ that exploit the fermionic (or bosonic) statistics of the trial state.

\subsection{Refined interactions}\label{sec.refine}
Our method for computing energies is based on a known fact: The second quantized Hamiltonian for fermions (bosons) in a single Landau level is invariant under any change of $H_L$ with even (odd) $L$. By contrast, the real-space potential Eq.~\eqref{eqn.hltovk} depends on \textit{all} $H_L$. Two different sets of real-space harmonics, $\{V_k\}$  and  $\{\tilde V_k\}$, are guaranteed to give the same energy Eq.~\eqref{eq.energy} for a fermionic (bosonic) state if their odd (even) Haldane pseudopotentials match. However, the statistical noise differs between the two sets. We exploit this freedom to greatly reduce numerical noise by choosing a suitable form of the harmonics $V_k$. From here on, we will focus on the fermionic case; for the bosonic one, `odd' and `even' are interchanged. 

Above, we identified the growth of $V_k$ at large $k$ as the source of numerical instability. Our resolution to this problem is to tune the even $H_L$ to suppress $V_k$ with large $k$. Given a real-space potential expressed through $V_k$, we find the corresponding Haldane pseudopotentials $H_L$. Then, we divide $H_L$ into odd $\vec{H}_\text{odd}=(H_1,H_3,\ldots)$ and  even $\vec{H}_\text{even}=(H_0,H_2,\ldots)$ parts written as vectors of dimensions $d_\text{odd}=\lfloor\frac{2q+1}{2}\rfloor$ and $d_\text{even}=\lfloor q \rfloor+1$, respectively. We now replace $\vec{H}_\text{even}$ with a vector $\tilde{\vec{H}}_\text{even}$ of free parameters and introduce the equivalent harmonics as
\begin{align}\label{eq.VkRefined}
    \tilde{\vec V} (\tilde{\vec{H}}_\text{even}) = N(q)\cdot \tilde{\vec H},
\end{align}
with the matrix $N(q)$ specified in Eq.~\eqref{eqn.nkl}. The vector $\tilde{\vec H} = \tilde{\vec{H}}_\text{even}\cup \vec{H}_\text{odd}$ is comprised of the original odd pseudopotentials $\vec{H}_\text{odd}$ and the free parameters $\tilde{\vec{H}}_\text{even}$. By construction, $    \tilde{\vec V}$ yields the same energy as ${\vec V}$.

To reduce statistical noise, we tune $\tilde{\vec{H}}_\text{even}$ so that $\tilde V_k$ decay quickly by minimizing the cost function 
\begin{align}\label{eq.opt}
    {\cal F}(\tilde{\vec{H}}_\text{even}) = \sum_{k=1}^{2q} [\tilde{V}_k(\tilde{\vec{H}}_\text{even})]^2 (2k+1)^\alpha,
\end{align}
with $\alpha\geq2$ to penalize large $k$ harmonics over small $k$.~\cite{footnote} We take $\alpha=4$ for the data presented in this work, but we find the final result to be largely insensitive of $\alpha$; see Fig.~\ref{fig.vk_opt}. After determining $\tilde{\vec{H}}^\text{opt}_\text{even}$ that minimizes ${\cal F}(\tilde{\vec{H}}_\text{even})$, we compute the refined values of the harmonics  $\tilde{V}_k(\tilde{\vec{H}}^\text{opt}_\text{even})$. Using them to compute the energy Eq.~\eqref{eq.energy} yields the same value as using the original $V_k=\tilde V_k({\vec{H}}_\text{even})$ but with less noise. The optimization procedure can be done semi-analytically, e.g., with Mathematica, and is not numerically expensive for any system size.

We note that Eq.~\eqref{eq.energy} assumes spherically symmetric states. Nevertheless, the refined harmonics $\tilde{\vec V}$ can be cast into a real-space form $\tilde{V}(\theta)$ with Eq.~\eqref{eq.expnasion2} that can be used to compute the energy of non-zero angular momentum states within Monte Carlo simulations.

\section{Laughlin states and CFL in higher Landau levels}
\label{sec.examples}

Before applying this technique to paired states, we demonstrate its efficacy for two simpler examples. For Laughlin states, we directly compare Monte Carlo energies to values obtained via exact diagonalization. For CFLs at half and quarter filling, we show that thermodynamic energy is insensitive to the shift, even though energies at moderate system sizes differ significantly.

\subsection{Laughlin state} 
\label{subsec.laughlin}

We compute the energy expectation value of the $\nu=\frac{1}{3}$ Laughlin state for Coulomb interactions in the 0-7LL with exact diagonalization and using the Monte Carlo approach discussed above. For the former, we use the DiagHam libraries\cite{DiagHam} to generate Laughlin states with up to $N_e=14$ particles based on Ref.~\onlinecite{Bernevig_Anatomy_2009}. Using Monte Carlo, we compute $G_k$ of the Laughlin state for various system sizes topping at $N_e=60$ electrons with $1.5 \times 10^9$ uncorrelated updates. Using Eq.~\eqref{eq.Vk$n$LL} we compute $V_k^\text{$n$LL}$ and refine the harmonics with procedure Eq.~\eqref{eq.VkRefined} to obtain $\tilde{V}^\text{$n$LL}_k$. Finally, we compute the energy Eq.~\eqref{eq.energy} with Monte Carlo from the density-density correlation function $G_k$ and plot it as a function of $N_e^{-1}$ in Fig.~\ref{fig.Laughlin_energy}; see Appendix~\ref{app.background} for our convention of normalizing energies. At system sizes where Monte Carlo and exact diagonalization are both applicable, the energies obtained with either approach match up to statistical errors. (For these system sizes, the error is too small to be visible in the scale of Fig.~\ref{fig.Laughlin_energy}.) The lowest Landau level energies and thermodynamic extrapolations reproduce values previously reported in the literature.~\cite{Ciftja_MC_2003,Jain_composite_2007}

To extract thermodynamic values of the trial energies, we include all data for system sizes $N_e \geq 10$. There is some ambiguity in the precise way how this limit is taken; this choice does not affect the asymptotic value at $N_e \rightarrow \infty$ but may result in stronger or weaker finite size effects. Our convention is described in Appendix~\ref{app.background}. For all eight Landau levels under consideration, a linear fit matches the data for $N_e \geq 10$ exceptionally well.

\begin{figure}
 \centering
\includegraphics[width=\linewidth]{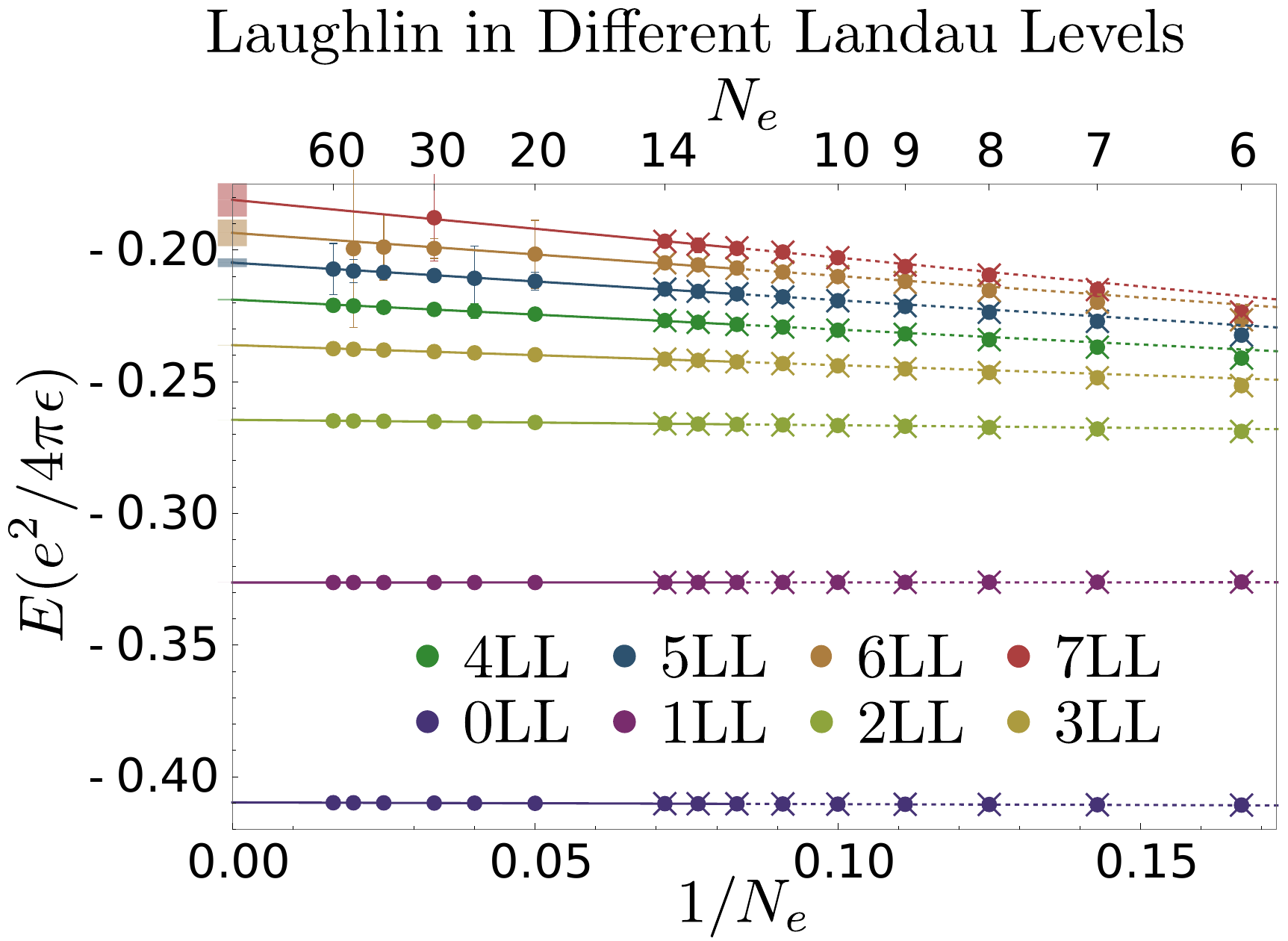}
 \caption{
The energy per particle of the Laughlin $\nu=\frac{1}{3}$ state is computed with Monte Carlo (circles) and exact diagonalization (crosses). The same Monte Carlo data, i.e., the set of numbers $G_k$, was used for all Landau levels. Convergence becomes worse for higher Landau levels. Still, we are able to reliably evaluate the Coulomb energy for large systems in as high as the seventh Landau level. The systems with $N_e\geq10$ are used for thermodynamic fit (solid line).}
  \label{fig.Laughlin_energy} 
 \end{figure}

\subsection{Composite Fermi liquid}
\label{subsec.cfl}

\begin{figure}
     \centering\includegraphics[width=\linewidth]{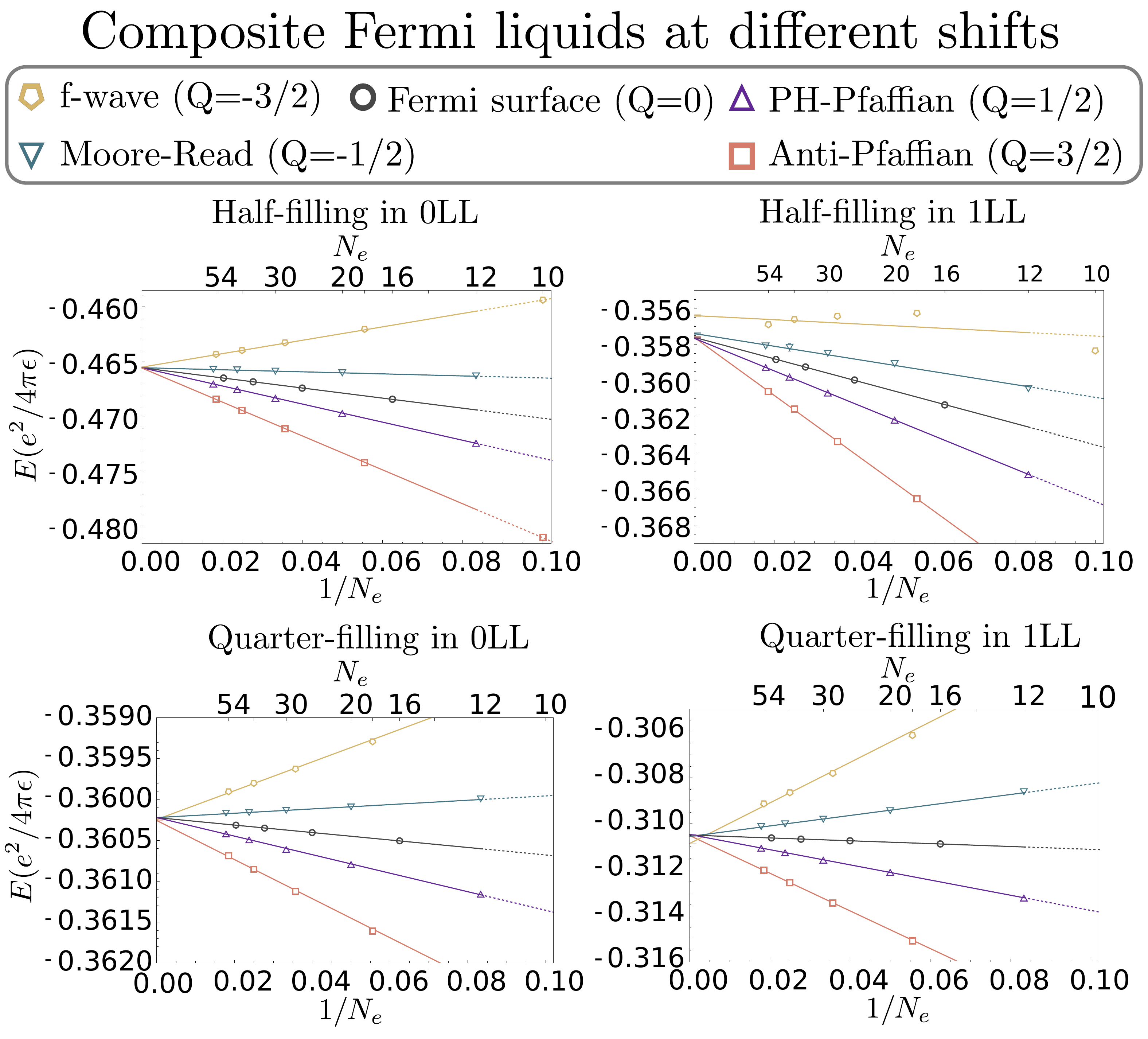}
     \caption{The energies of composite Fermi liquids in 0LL and 1LL at half and quarter fillings for different shifts. Their wavefunctions are defined in different Hilbert spaces determined by the strength $Q$ of the effective monopole experienced by the composite fermions. The finite size energies depend strongly and systematically on the shift. Nevertheless, the thermodynamic energies agree remarkably well. }
     \label{fig.CFL}
 \end{figure}

Composite Fermi liquids are metallic states that compete with paired states at half and quarter filling. Trial wavefunctions for such states can be obtained in a planar geometry by attaching the appropriate number $2p$ of flux quanta to a Slater determinant of plane-wave states,\cite{Rezayi_fermi-liquid-like_1994} i.e.,
\begin{align}
\label{eqn.cfl}
\Psi_\text{CFL} = P_\text{LLL} \det[e^{i \vect k_j \cdot \vect r_i}]\prod_{i<j}(z_i -z_j)^{2p}. 
\end{align}
Here, we omitted the Gaussian factor and denoted projection into the lowest Landau level by $P_\text{LLL}$. In the spherical geometry used for all our numerics, the differences $z_i-z_j$ are replaced by $u_iv_j-u_jv_i$ with $u_i=\cos\frac{\theta_i}{2}e^{i\frac{\phi_i}{2}}$ and $v_i=\sin\frac{\theta_i}{2}e^{-i\frac{\phi_i}{2}}$. The plane-wave states in the determinant become monopole harmonics $e^{i \vect k_j \cdot \vect r_i}~\to~Y^{Q}_{l_i,m_i}(\vect r_j)$. These harmonics are parameterized by the total angular momentum $l_i$, the $z$-projection $m_i$, and an integer or half-integer parameter $Q$ encoding the possible boundary conditions of single-particle states on a sphere. Its meaning is the effective monopole strength experienced by the composite fermions, which is related to the \textit{shift}
\begin{align}
{\cal S} = \nu^{-1}N_e - 2 q
\end{align}
via ${\cal S}=2(p -Q)$. Rotationally symmetric states occur at particle numbers $N_e = (l_F+1)^2-Q^2$, where composite fermions fill the lowest $n_F \equiv l_F-|Q|$ Landau levels. An equivalent form of wavefunction $\Psi_\text{CFL}$ is
\begin{align}
\Psi_\text{CFL} = P_\text{LLL}\Phi_{n_F} (\Phi_1)^{2p}~,
\end{align}
where $\Phi_n$ denotes the Slater determinant of $n$ filled Landau levels. The number of filled Landau levels depends on the particle number via $N_e=n_F(n_F+2Q)$. As usual in Monte Carlo studies of quantum Hall wavefunctions, we employ a modified form of projection into the lowest Landau level to enable efficient simulations (cf.~Appendix~\ref{app.projection}).

These wavefunctions for different shifts all represent the same CFL phase. Their energies at finite $N_e$ depend strongly and systematically on the shift. Still, the thermodynamic energies extracted from linear fits for different $Q$ coincide. In Fig.~\ref{fig.CFL}, we show the thermodynamic extrapolations of 0LL and 1LL Coulomb energies for the half-filled and quarter-filled CFLs with different shifts. In Appendix~\ref{app.background}, we further show that the effects of the shift can be systematically canceled by introducing a single interaction-dependent parameter that results in an almost perfect collapse of the data. 

We note the importance of taking these effects into account. Comparing the energy of an even-denominator state to a CFL at a different shift leads to systematic errors at finite size.

\section{Paired states} 
\label{sec.paired}
Within composite fermion theory, the even-denominator states at $\nu=\frac{1}{2p}$ can be viewed as a superfluid of composite fermions with $2p$ flux quanta.~\cite{Read_paired_2000} This description can be used to deduce many properties of the states, such as their topological order, edge structure, and physical observables. In particular, different pairing channels give rise to distinct topological orders. Consequently, determining the favored pairing channel is an important question to study both experimentally and numerically.

To obtain a trial wavefunction for paired states, one replaces the Slater determinant in Eq.~\eqref{eqn.cfl} with a Pfaffian factor. For an arbitrary odd pairing channel $\ell=2r+1$, we take~\cite{Yutushui_Large_scale_2020}
\begin{align}
\label{eqn.apr}
\Psi_\ell = P_\text{LLL}\text{Pf}\left(
    \frac{1}{z_{ij}}
    \left[\frac{z^*_{ij}}{z_{ij}}\right]^{r}\right)\prod_{i<j}(z_i -z_j)^{2p} ~.
\end{align}
For $\ell=1$, this wavefunction reduces to the celebrated Moore-Read Pfaffian \cite{Moore_nonabelions_1991}, which does not require any explicit Landau-level projection. The anti-Pfaffian and f-wave topological orders correspond to $\ell=-3$ and $\ell=3$, respectively. For these two phases, different trial states based on the `parton' framework are also available.~\cite{Jain_Incompressible_1989,Wen_Non_Abelian_1991} Their wavefunctions can be generated from $SU(2)_{\pm 2}$ chiral algebra~\cite{Henderson_CFT_2024} and take the form 
\begin{align}
\label{eqn.parton}
\Psi_{SU(2)_{\pm 2}} = P_\text{LLL} \Phi_{\pm 2}^2 \Phi_1^{2p \mp 1}~,
\end{align}
where $\Phi_{-2}\equiv \Phi^*_2$.~\cite{Wen_Non_Abelian_1991} The wavefunctions $\Psi_{SU(2)_{\pm 2}}$ differ microscopically from $\Psi_{\ell=\pm 3}$ but describe the same topological orders.~\cite{Balram_parton_2018} The energies of parton and composite-fermion-based wavefunctions can differ, even after thermodynamic extrapolation, and we thus include both in our study. At half filling, the state $\Psi_{SU(2)_{2}}$ cannot be efficiently projected to the lowest Landau level (cf. Appendix~\ref{app.projection}), and we use this state only at quarter filling.

\begin{figure}
    \centering
    \includegraphics[width=\linewidth]{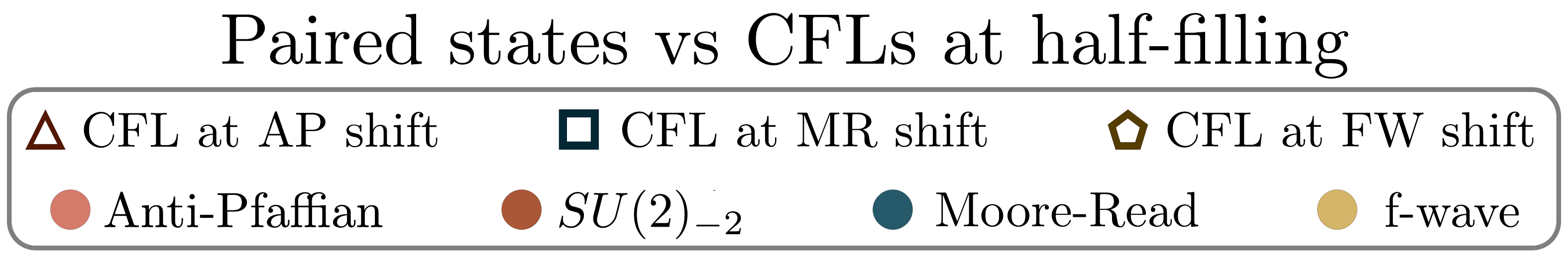}\\ 
    \includegraphics[width=0.49\linewidth]{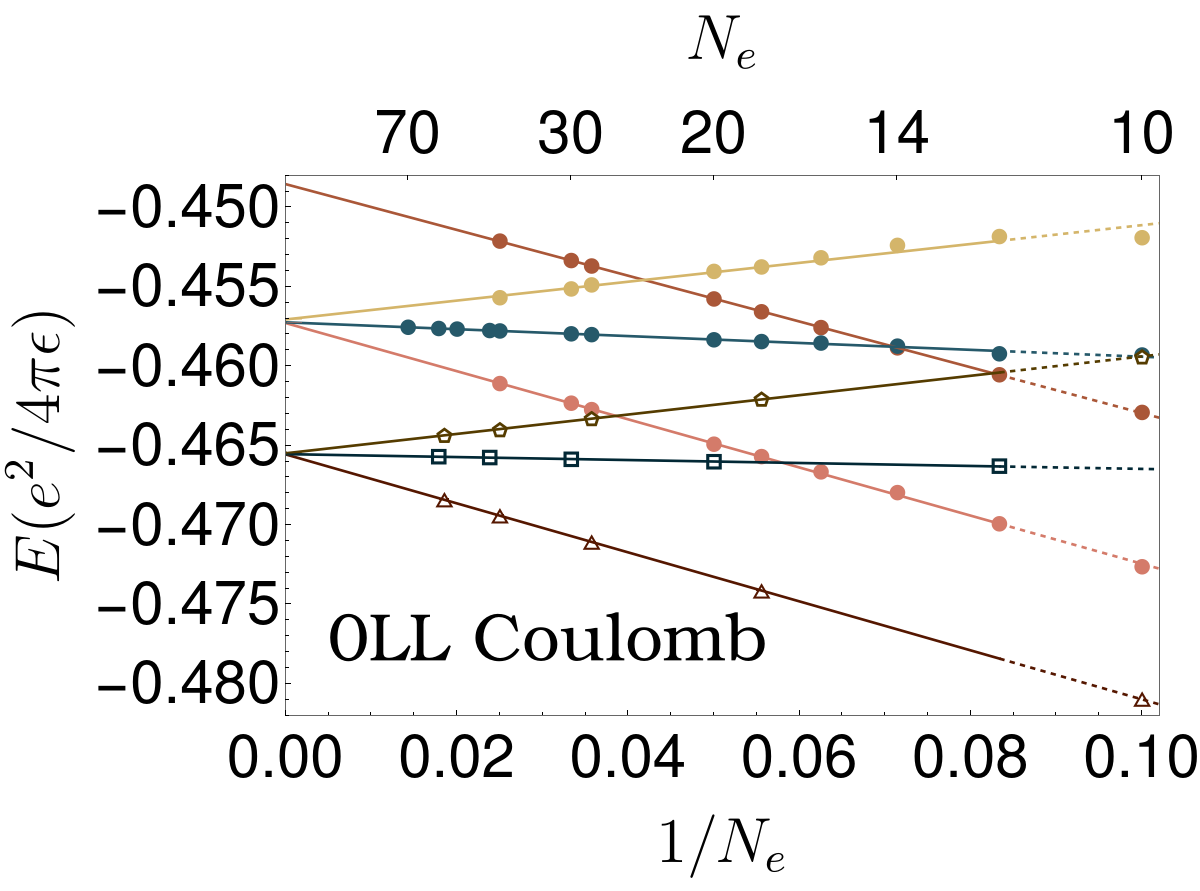}
    \includegraphics[width=0.49\linewidth]{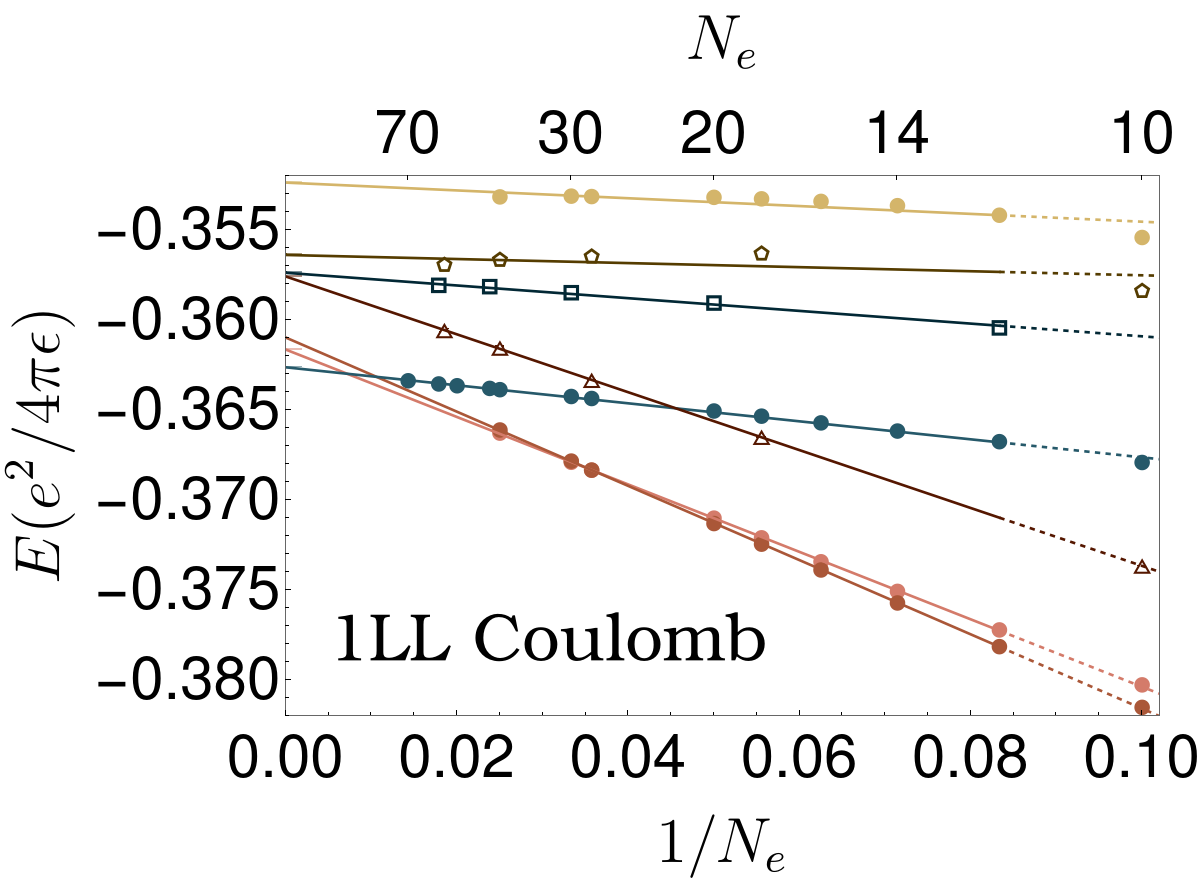}
    \caption{The energies of 0LL and 1LL Coulomb interactions for various trial states at half filling. In the 0LL, the composite Fermi liquid is favored, while in 1LL, the Moore-Read or anti-Pfaffian phases are favored. The energies of Moore-Read and anti-Pfaffian do not coincide since the trial wavefunctions representing these phases are not exactly particle-hole conjugates of each other.}
    \label{fig.Coul_Half}
\end{figure}

\subsection{Half-filled states} 
The half-filled Landau level is the most extensively investigated even denominator; see Ref.~\onlinecite{Ma_Fractional_5_2_2022} for a recent review. Extensive numerical studies and experiments in many platforms, e.g., GaAs or graphene, agree that 0LLs with Coulomb favor a compressible CFL at this filling, while 1LL favors incompressible states. The nature of the latter remains the subject of an ongoing debate. Its resolution requires confronting a special feature of the half-filled Landau level compared to other even-denominators: Particle-hole symmetry ${\cal C}: \nu\rightarrow 1-\nu$ relates different competing phases. They are consequently degenerate for any two-body interactions in a single Landau level and considerable effort has been devoted to distinguish them in the presence of Landau-level mixing.

The filling factor $\nu=\frac{1}{2}$ is well suited for numerical works with particle numbers  $N_e=2,4,\ldots,20$ readily accessible. Numerical studies based on exact diagonalization have found that the CFL in the 0LL quickly gives way to Moore-Read or anti-Pfaffian state upon tuning $H_1$.~\cite{Rezayi_incompressible_2000,moller_paired_2008,Wojs_landau_level_2010} More recent work suggests that modifying the Coulomb interactions of the third Landau level in monolayer graphene by $H_1$ or $H_3$ can induce a state with f-wave pairing.~\cite{Kim_Even_Denominator_f_wave_2019,Sharma_unconventional_2022}

\begin{figure}
 \centering
\includegraphics[width=0.99\linewidth]{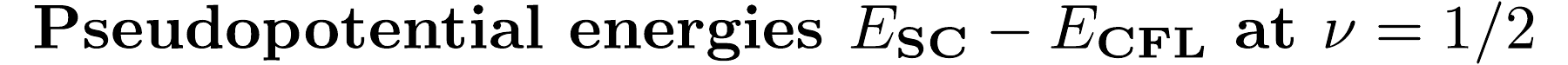}\\
\includegraphics[height=0.62\linewidth]{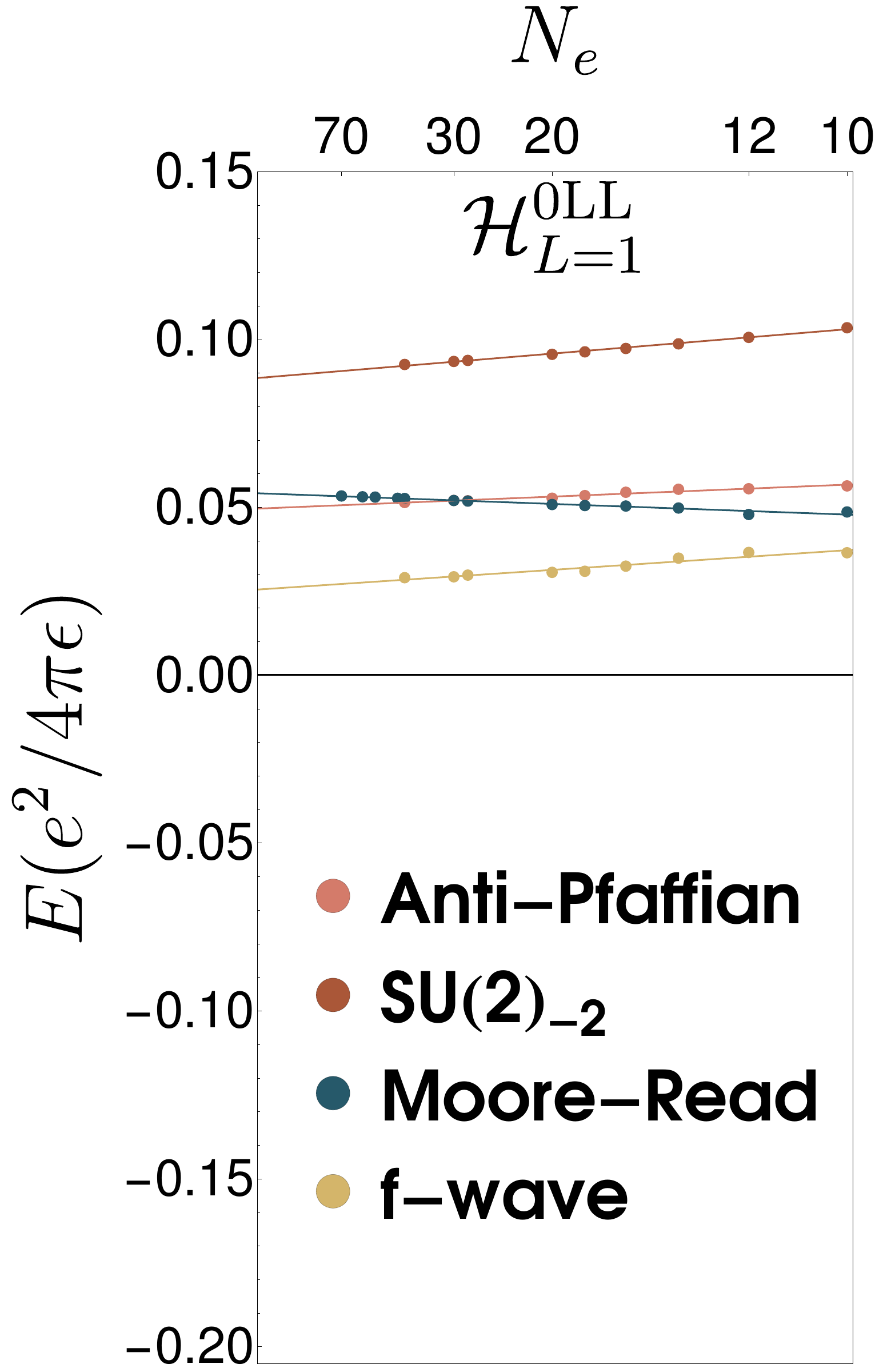}
\includegraphics[height=0.62\linewidth]{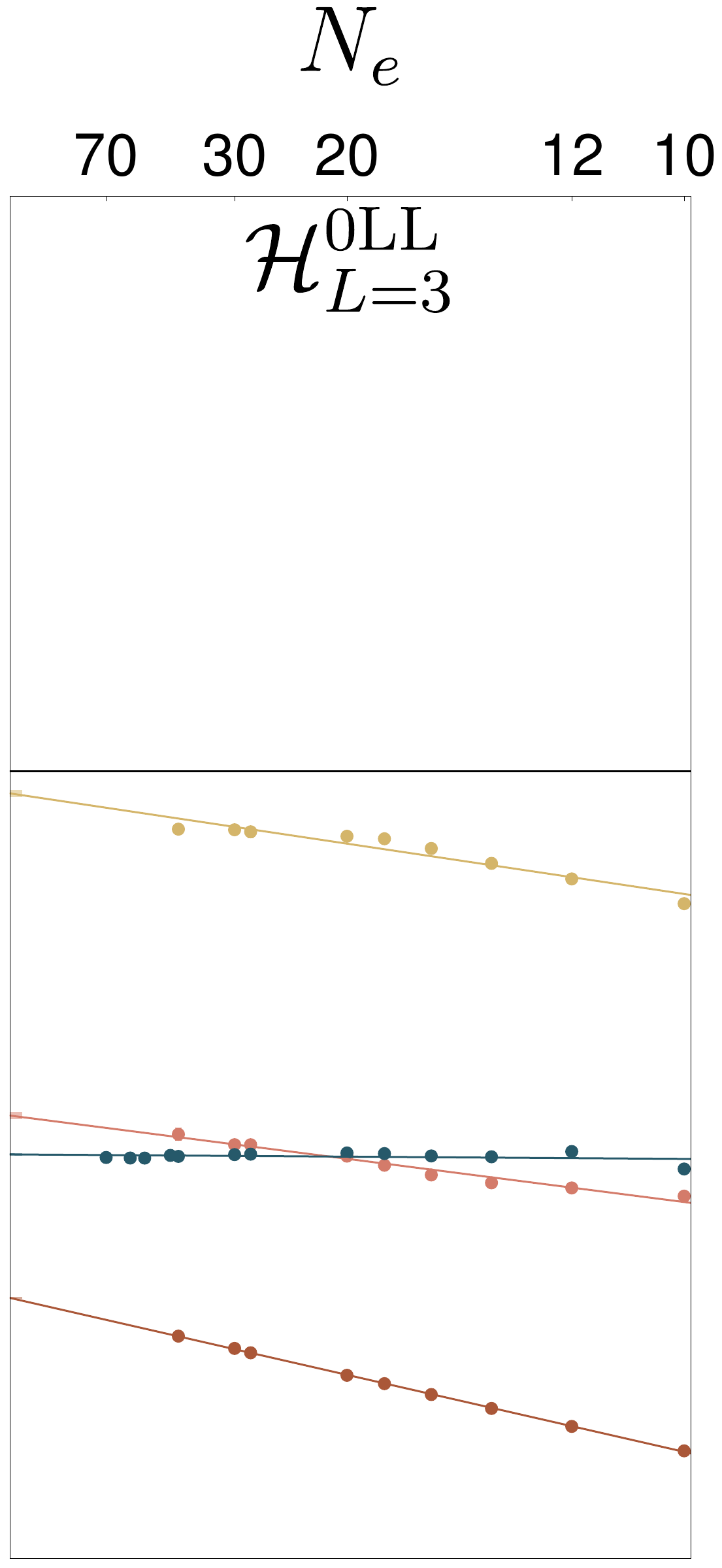}
\includegraphics[height=0.62\linewidth]{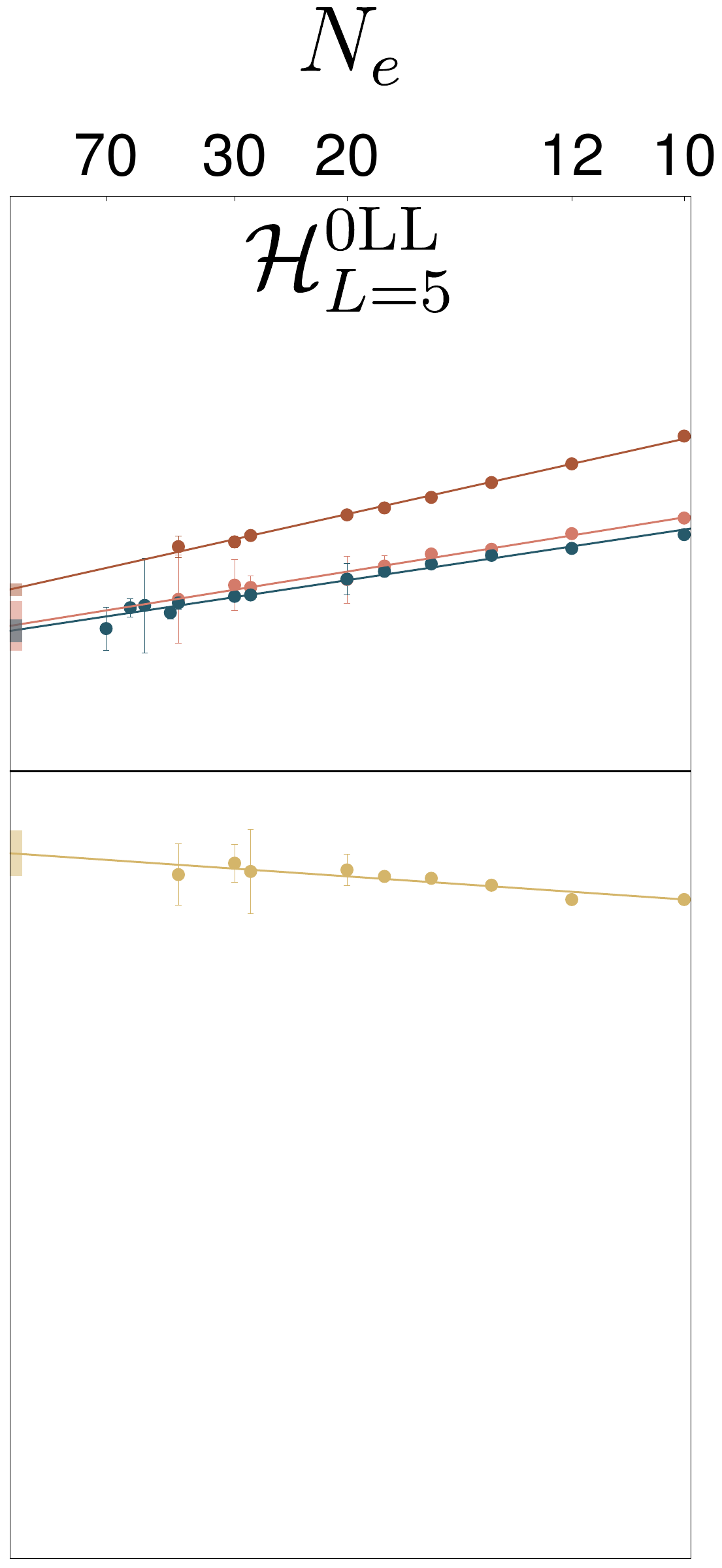}
 \caption{The energy difference between paired states and composite Fermi liquids of the Hamiltonian consisting of a single Haldane pseudopotential, ${\cal H}^\text{0LL}_L = H^\text{0LL}_L\hat{\cal P}_L$. The energies are computed via Monte Carlo from Eq.~\eqref{eq.energy} with refined harmonics. }
 \label{fig.Haldane_half} 
 \end{figure} 

In our numerical studies at half filling, we included trial states for the three main pairing channels in addition to the CFL. Thermal transport measurements in GaAs indicate a fourth topological order known as the particle-hole symmetric Pfaffian\cite{Son_is_2015}. However, no trial state for this phase in the lowest Landau level is known.\cite{Mishmash_numerical_2018,Yutushui_Large_scale_2020,Rezayi_Stability_2021}

{\it Coulomb energies.---}Firstly, we compute the energies of the trial states for Coulomb interactions in 0LL and 1LL. Fig.~\ref{fig.Coul_Half} shows the energies of the various trial states for system sizes up to $N_e=56$ ($N_e=70$ for the Moore-Read state, where explicit projection is not required). The data for all but the smallest systems are well described by linear fits to $N_e^{-1}$  (cf.~Appendix~\ref{app.background} for details). In the 0LL, CFL is favored over all pairing channels, in agreement with previous works. By contrast, in 1LL, the Moore-Read and anti-Pfaffian wavefunctions have lower energies than the CFL. The linear fits for the Moore-Read and the anti-Pfaffian trial states extrapolate to somewhat different thermodynamic values. This discrepancy does not pose any contradiction to particle-hole symmetry. The specific trial states we use for the Moore-Read or anti-Pfaffian phases are \textit{not} exact hole-conjugate of each other. Instead, ${\cal C} \Psi^{\nu=1/2}_{\ell=-3(1)} $ should be viewed as an alternative trial state for the Moore-Read (anti-Pfaffian) phase, which is microscopically distinct from $\Psi^{\nu=1/2}_{\ell=1(-3)}$. As a nonuniversal quantity, the energy is determined by microscopic properties and not only by the topological phase.~\cite{Henderson_Energy_2023,Sharma_BCS_2021}

{\it Haldane pseudopotentials.---}Each paired state directly competes with a CFL at the same shift for $N_e$ such that an integer number of composite-fermion levels are filled. To understand which interactions favor paired states, we compute the energy difference between the paired states and CFLs for the first three pseudopotentials, i.e., ${\cal H}_L\equiv H_L\hat{\cal P}_{L}$ with $L=1,3$ and $5$, which are shown in Fig.~\ref{fig.Haldane_half} (it becomes increasingly more complicated to reliably extract the energy for individual pseudopotentials with larger $L$; see Appendix~\ref{app.higher_pot}). We find that CFL has the lowest energy for ${\cal H}_1$, the Moore-Read or anti-Pfaffian for ${\cal H}_3$, and f-wave for ${\cal H}_5$. Knowing that CFL is favored in the lowest Landau level, these results imply that reducing $H_1$ or increasing $H_3$ will drive a transition into an incompressible state with the Moore-Read or anti-Pfaffian pairing. The relative increase of a third over the first pseudopotential is also the primary difference between 0LL and 1LL Coulomb interactions. The f-wave state can instead by targeted by the $L=5$ pseudopotential, which is naturally prominent in the third Landau level of monolayer graphene.~\cite{Kim_Even_Denominator_f_wave_2019}

\begin{figure}
    \centering
    \includegraphics[width=\linewidth]{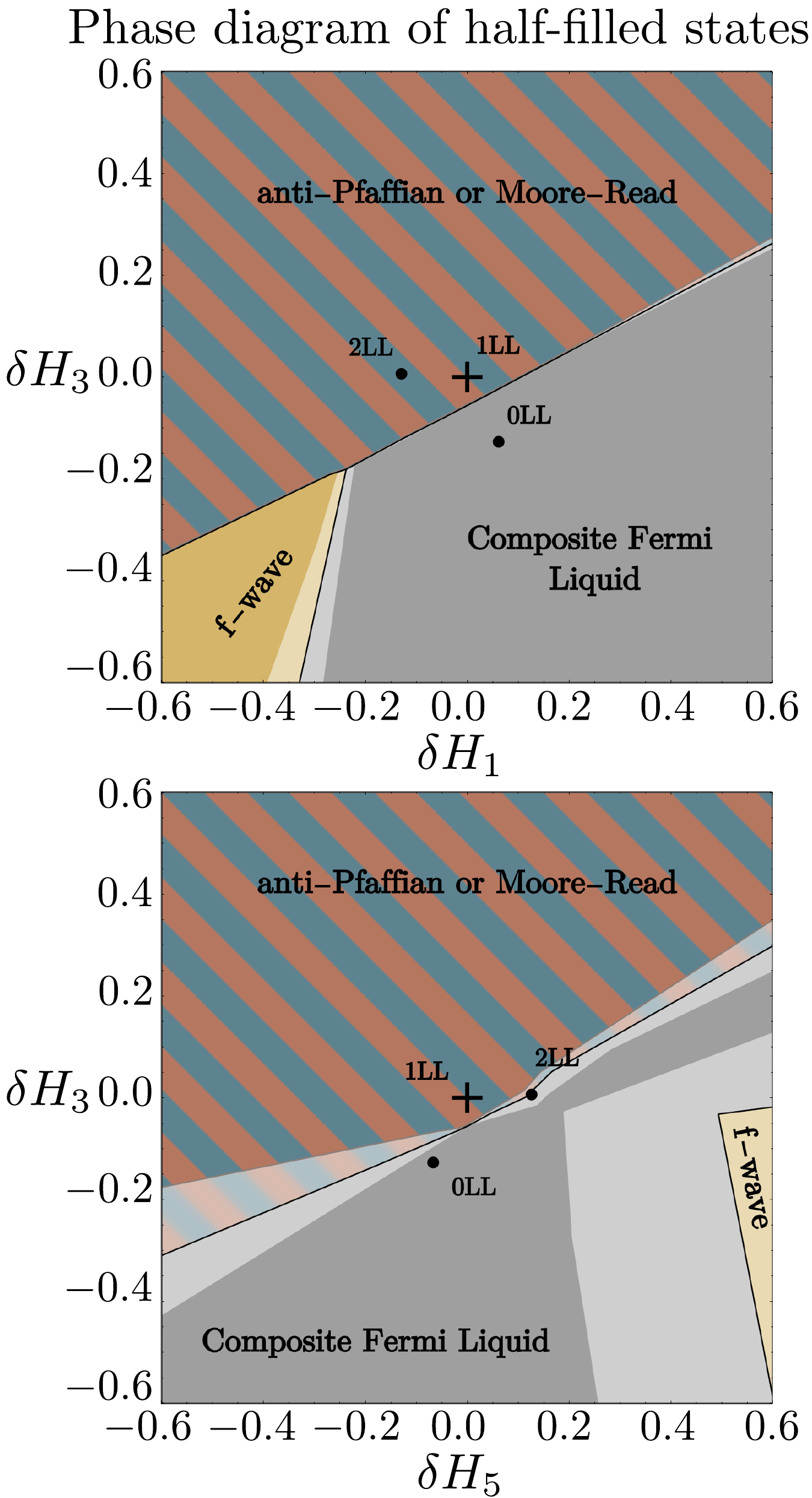}
    \caption{The phase diagram of half-filled states around 1LL Coulomb (denoted by $\boldsymbol{+}$) perturbed by $\delta H_1$, $\delta H_3$, and $\delta H_5$ was obtained by comparing the energies of the trial states. As reference points, we mark the pseudopotential values of 0LL and 2LL Coulomb Hamiltonians. The anti-Pfaffian and Moore-Read are favored by 1LL Coulomb and give way to composite Fermi liquid when $H_1$ and $H_5$ are increased, or $H_3$ decreased. }
    \label{fig.phase_half}
\end{figure}

{\it Phase diagram.---}Finally, we construct the phase diagram by comparing the energies of all trial states in the vicinity of 1LL Coulomb interactions. To this end, we extract the thermodynamic energies for 1LL Coulomb interactions, $E^{1LL}(\Psi)$, and of individual Haldane pseudopotentials, $E_L(\Psi)$. We then compare the energies
\begin{align}
    E(\delta H,\Psi) = E^{1LL}(\Psi) + \sum_{L=1,3,5} \delta H_L E_L(\Psi),
\end{align}
where $\delta H_L$ is measured in units of $H^{1LL}_L$. The resulting phase diagram in the $H_{L=1}$--$H_{L=3}$ and $H_{L=3}$--$H_{L=5}$ planes is shown in Fig.~\ref{fig.phase_half}. The CFL is stabilized in the vicinity of 0LL, while Moore-Read and anti-Pfaffian are favored near the 1LL point. The f-wave pairing channel has lower energies for strong negative deviations of $H_1$ and $H_3$ and positive $H_5$.

This phase diagram confirms qualitative features of tuning $\delta H_1$ and $\delta H_3$ obtained from exact diagonalization,\cite{Rezayi_incompressible_2000,Wojs_landau_level_2010} i.e., positive $\delta H_1$ and negative $\delta H_3$ derive system into gapless CFL phase.

\subsection{Quarter-filled states}
\label{subsec.quarter}

At quarter filling, the Hilbert space size at the same particle number is significantly larger than at half filling. Consequently, exact diagonalization studies are limited to relatively small system sizes with $N_e\leq12$, and Monte Carlo simulations play an important role in their study. Motivated by the experimental observation of quarter-filled states in wide quantum wells,~\cite{luhman_Observation_2008,shabani_evidence_2009} numerical studies focused on finite-width effects.~\cite{Papic_quarter_2009,Papic_interaction_2009} These works found that the simplest model of the finite-width potentials~\cite{Zhang_width_1986} can drive the transition from CFL to the incompressible Moore-Read or Halperin (553) state. Later, other paired states were studied at quarter filling, and the f-wave state was found to be energetically favored by a realistic charge distribution in a wide quantum well.~\cite{Faugno_Prediction_2019,Sharma_CF_pairing_quart_2024} In our work, we systematically study generic interactions via their trial-state energies. Our findings can be readily applied to other systems where paired states were observed, such as hole-doped GaAs~\cite{Wang_even_3_4_2022,Wang_quarter_hole_2023} and bilayer graphene.~\cite{Kumar_Quarter_2024} In particular, the main results in Fig.~\ref{fig.phase_quart} can be viewed as perturbations of the $N=0$ orbital of bilayer graphene since the interactions there are the same as in the 0LL of narrow GaAs quantum wells.
\begin{figure}
    \centering
    \includegraphics[width=\linewidth]{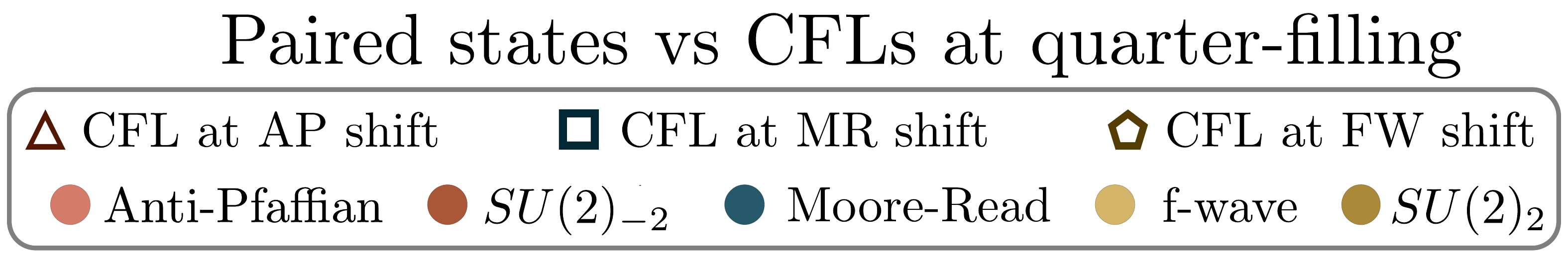}\\
    \includegraphics[width=0.49\linewidth]{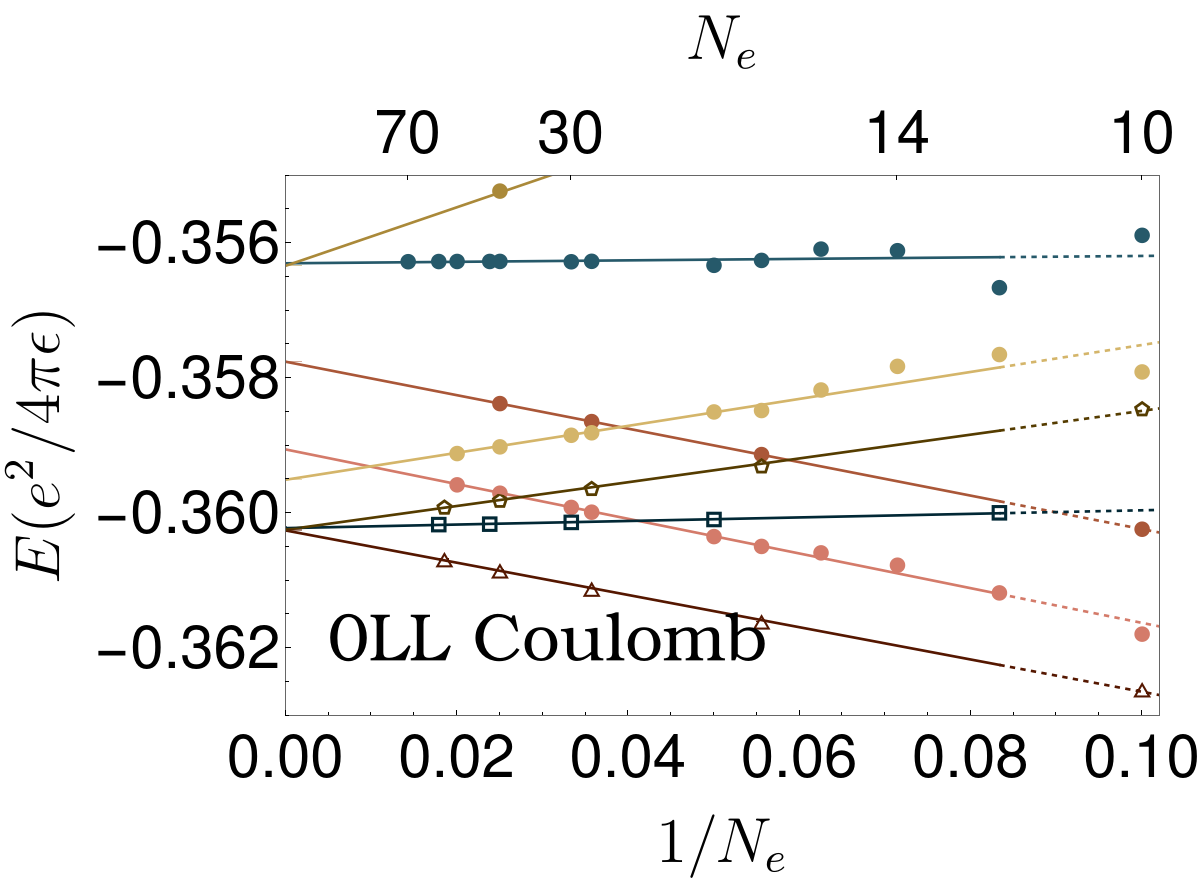}
    \includegraphics[width=0.49\linewidth]{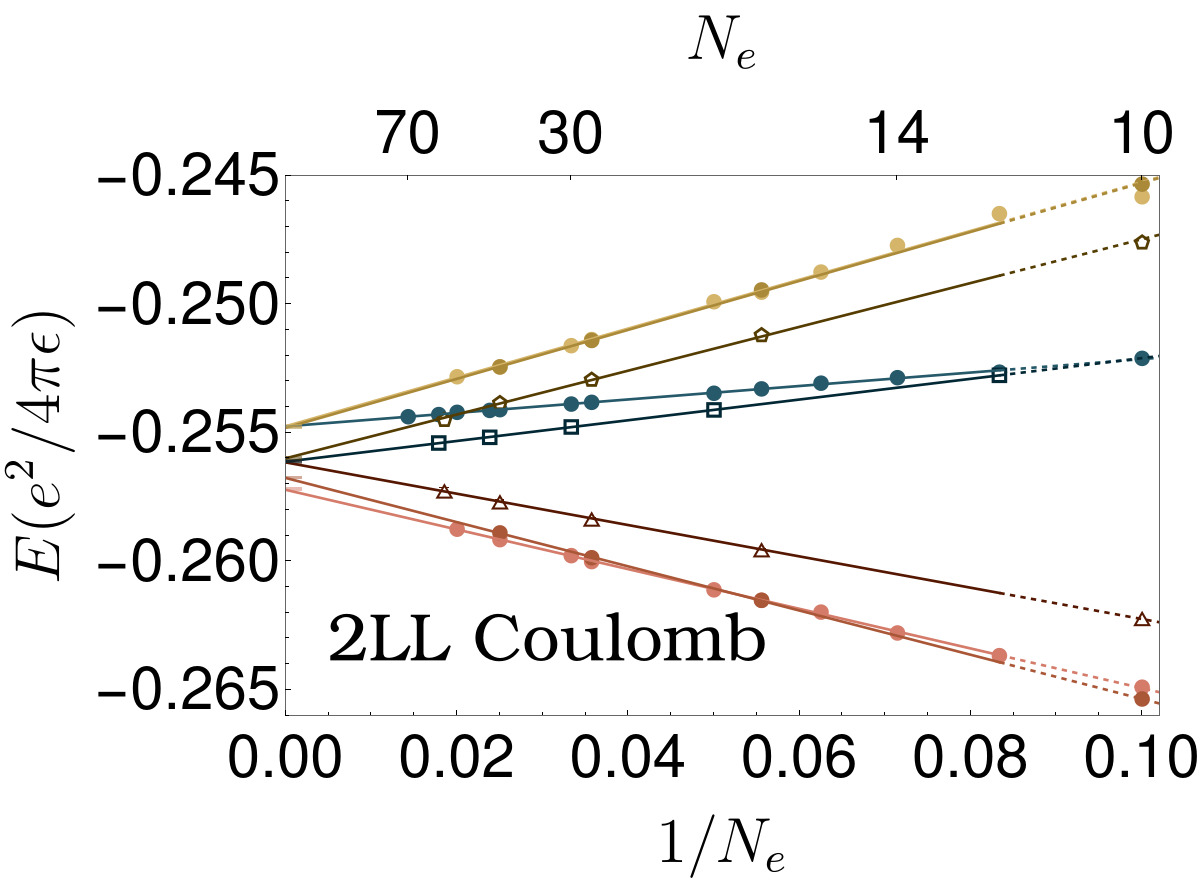}
    \caption{Quarter-filled states in 0LL and 2LL. In the 0LL, the composite Fermi liquid has the lowest energy. The Moore-Read and anti-Pfaffian states have different energies, and the latter is favored in 2LL. }
    \label{fig.Coul_Quart}
\end{figure}

At quarter filling, we study the energy of the trial states Eqs.~\eqref{eqn.cfl}, \eqref{eqn.apr}, and \eqref{eqn.parton} with $p=2$. In contrast to half filling, the Moore-Read and anti-Pfaffian phases are not particle-hole conjugates of each other and have different energies for two-body interactions.

{\it Coulomb energies.---}We start by computing thermodynamic energies of quarter-filled states for Coulomb interactions in the three lowest Landau levels. We find that in the 0LL and 1LL, the CFL is favored, while in 2LL, the anti-Pfaffian state has the lowest energy; see Fig.~\ref{fig.Coul_Quart}. 

\begin{figure}
 \centering
\includegraphics[width=0.99\linewidth]{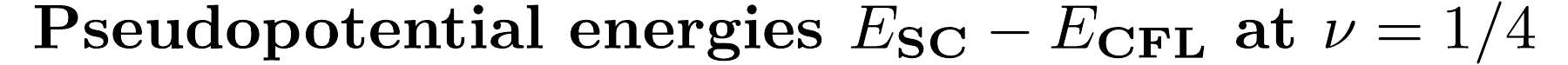}\\
\includegraphics[height=0.62\linewidth]{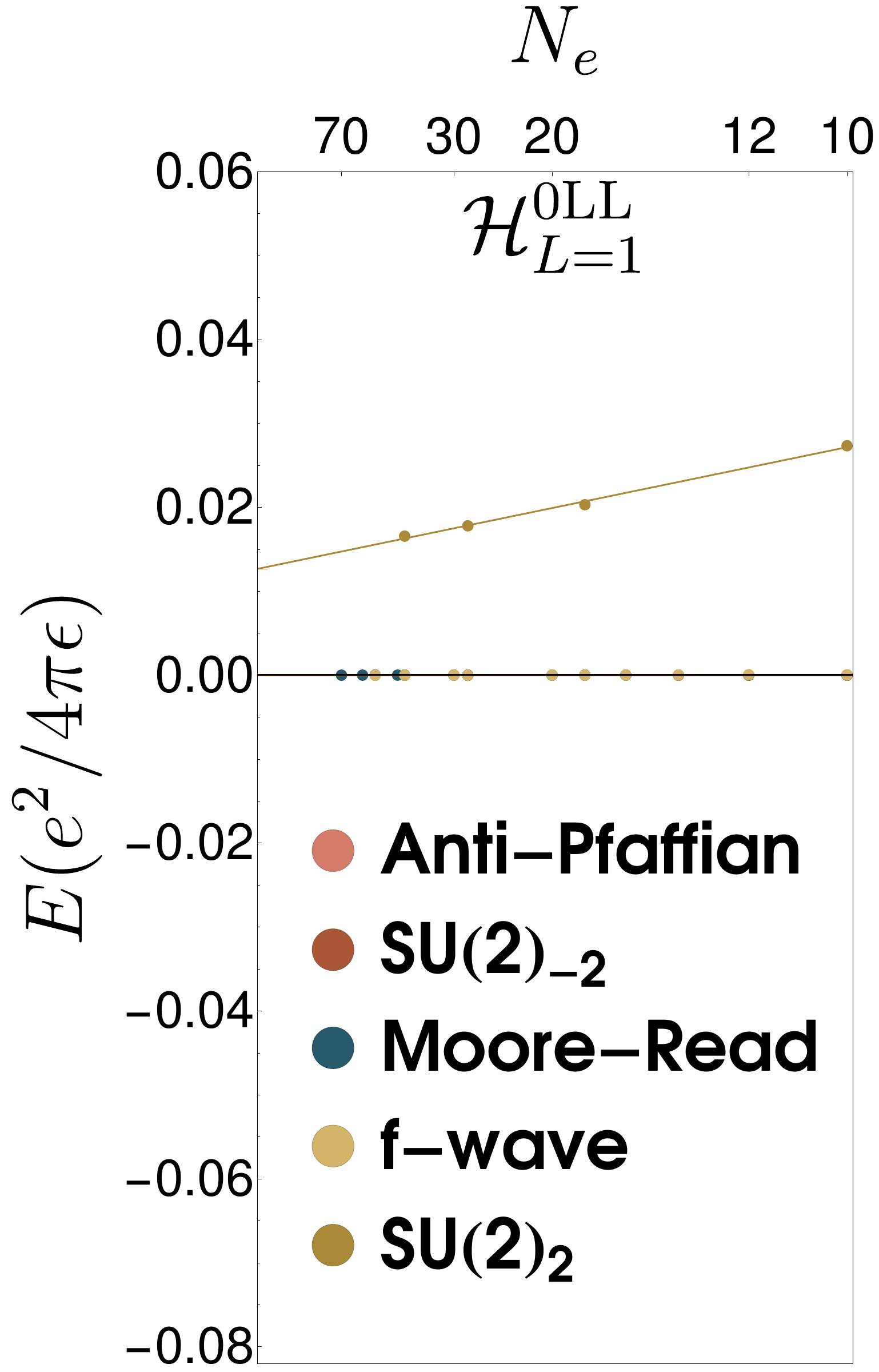}
\includegraphics[height=0.62\linewidth]{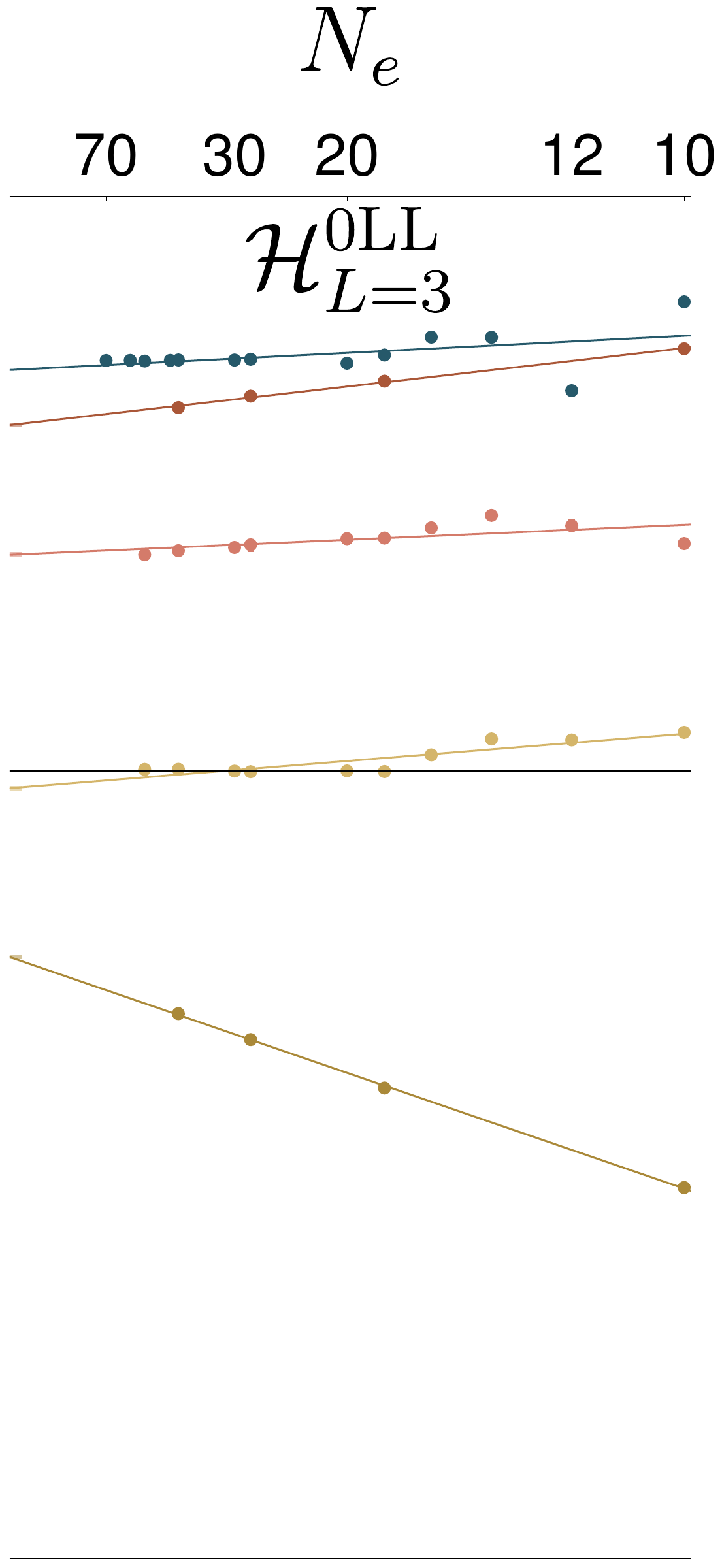}
\includegraphics[height=0.62\linewidth]{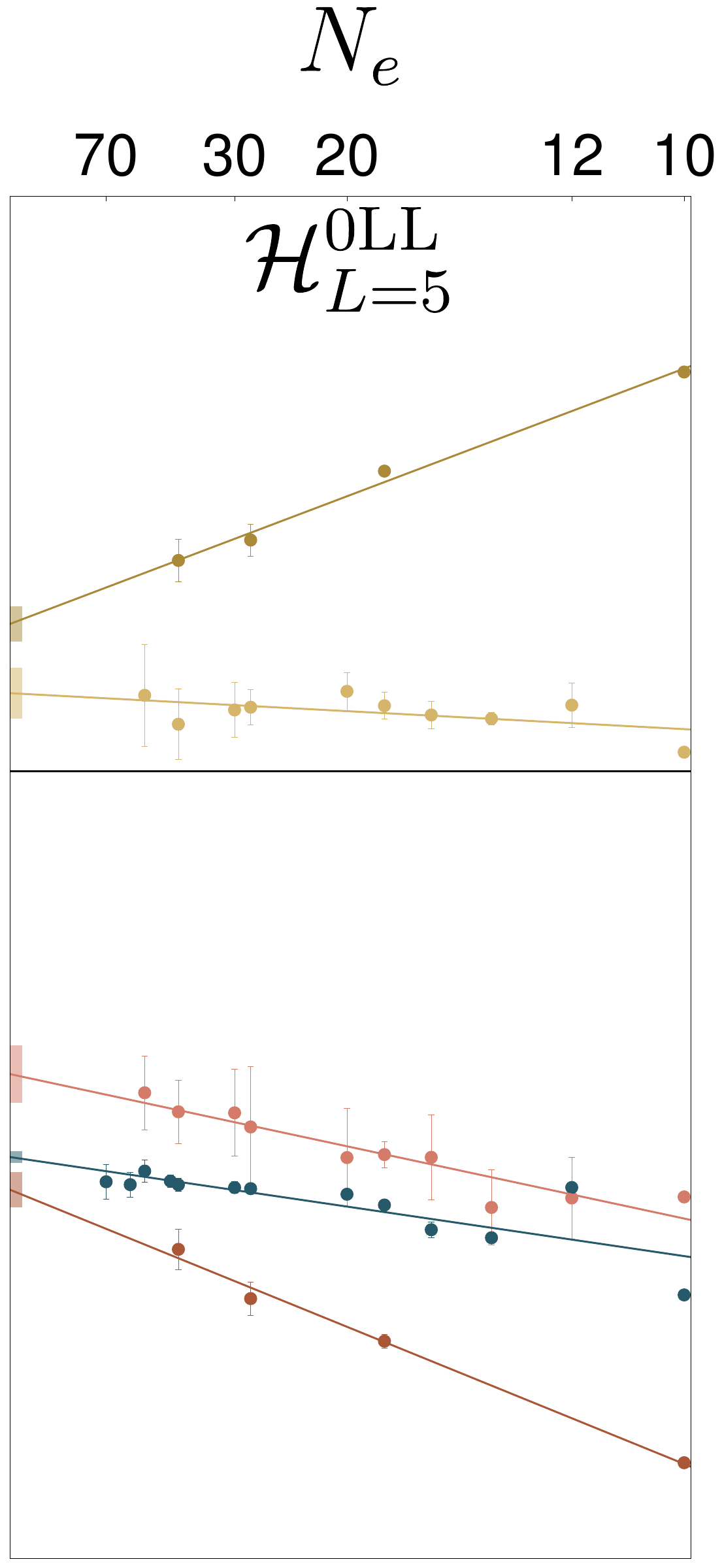}
 \caption{The energy difference between paired states and composite Fermi liquids for a Hamiltonian consisting of a single pseudopotential at 0LL Coulomb value, ${\cal H} = H^\text{0LL}_L\hat{\cal P}_L$. The energies are computed via Monte Carlo from Eq.~\eqref{eq.energy} with refined harmonics. }
 \label{fig.Haldane_quart} 
 \end{figure}

{\it Haldane pseudopotentials.---}We proceed by computing the thermodynamic energies of $L=1,3,5$ Haldane pseudopotentials; see Fig.~\ref{fig.Haldane_quart}. The projection method we use renders $E_{L=1}=0$ for all trial states except $\Psi_{SU(2)_2}$. The large difference between energies of CFL and paired state for ${\cal H}_{L=3}$ and ${\cal H}_{L=5}$ suggests that these Haldane pseudopotentials can be used to tune between the phases. The f-wave state has opposite behavior compared to Moore-Read and anti-Pfaffian, i.e., positive $\delta H_3$ and negative $\delta H_5$ favor f-wave over CFL.

{\it Phase diagram.---}Similarly to the half-filled state, we compute the phase diagram of trial states Fig.~\ref{fig.phase_quart} by perturbing 0LL Coulomb with $\delta H_3$ and $\delta H_5$. We find that any of the three main pairings is favored in certain regions of the phase diagram. The anti-Pfaffian pairing requires the smallest modification of the 0LL Coulomb to reach the paired state.

{\it Exact diagonalization.---}Finally, we corroborate the phase diagram Fig.~\ref{fig.Haldane_quart} using exact diagonalization for $N_e=10$ particles. We use Monte Carlo sampling to obtain a Fock-space wavefunction by computing overlaps with the $10^6-10^7$ basis states (depending on the shift). To suppress statistical noise, we project the Fock-space wavefunction onto zero angular momentum as in Ref.~\onlinecite{Mishmash_numerical_2018}.

We chose a direction $\theta_\Psi$ (indicated by arrows in Fig.~\ref{fig.phase_quart}) in the $\delta H_3-\delta H_5$ plane that favors the paired state $\Psi$. We compute the overlaps of the trial states with the exact ground state of 
\begin{align}
    {\cal H} = {\cal H}^\text{0LL} + x({\cal H}^\text{0LL}_3 \cos\theta_\Psi + {\cal H}^\text{0LL}_5 \sin\theta_\Psi)\label{eqn:hx}
\end{align}
as a function of $x$, see Fig.~\ref{fig.overlaps}. For the anti-Pfaffian and f-wave pairings, where the CFL wavefunctions Eq.~\eqref{eqn.cfl} are defined for $N_e=10$, we observe a transition at a positive value of $x$ for which the paired states exhibit higher overlap with the ground state. The largest overlaps with the exact-diagonalization ground states are significantly larger than the ones between paired states and CFLs, see Appendix~\ref{app.overlaps}. Meanwhile, negative $x$ favors CFL, which is reflected in an increasing overlap. We note that the transitions for small systems occur closer to the Coulomb point than in the thermodynamic limit. This behavior is consistent with Fig.~\ref{fig.Haldane_quart}, which shows larger energy differences between trial states at smaller particle numbers.

At the Moore-Read shift, $N_e=10$ does not correspond to a filled shell, and the CFL has non-zero angular momentum. The ground state in the vicinity of 0LL also has non-zero angular momentum, which is consistent with it being a CFL. At a positive value of $x$, the ground state becomes a zero angular momentum state with substantial overlap with the Moore-Read wavefunctions. The overlap squared with Moore-Read reaches $0.97$ at $x=0.075$ and then decays.

\section{Discussion}

In this work, we developed a method of computing energies of quantum Hall trial wavefunctions for large systems with generic interactions. We used it to extract the {\it thermodynamic} energies of the first three Haldane pseudopotentials for different pairing channels of even-denominator states and their direct competitors, the CFLs. This allows us to construct the phase diagrams, Figs.~\ref{fig.phase_quart} and \ref{fig.phase_half}, based on the energies of available trial states. We corroborated the phase diagram at quarter filling with exact diagonalization results; see Fig.~\ref{fig.overlaps}. The results for half filling are consistent with earlier studies, e.g., Ref.~\onlinecite{Rezayi_incompressible_2000,Wojs_landau_level_2010}. Still, we caution that our analysis does not allow for other competing phases, such as Wigner crystals, which frequently occur at low filling factors. Ultimately, the direct identification of pairing channel can be achieved through thermal transport experiments;~\cite{Banerjee_observation_2018,Dutta_Distinguishing_2022,Dutta_Isolated_2022,Melcer_Absent_2022,Paul_Thermal_2023,Melcer_Heat_2024,Paul_Thermal_2024,Dutta_novel_2022} see also other proposals.~\cite{Manna_Full_Classification_2022,Yutushui_Identifying_2022,Park_Fingerprints_2024,Yutushui_Identifying_2023,Yutushui_Localization_2024,Yutushui_universal_2024}

We find that the anti-Pfaffian paired state at quarter filling requires the smallest modification of the 0LL Coulomb Hamiltonian. Since the two-body interactions are particle-hole symmetric in a single Landau level, our findings equally apply to $\nu=\frac{1}{4}$ and $\nu=\frac{3}{4}$.

A recent experiment Ref.~\onlinecite{Kumar_Quarter_2024} observed an incompressible state at $\nu=\frac{3}{4}$ in several $N=0$ orbitals of bilayer graphene. By contrast, compressible states were observed at $\nu=\frac{1}{4}$ in the same Landau levels, which indicates a significant breaking of particle-hole symmetry, i.e., Landau-level mixing.  Additional interactions arising from Landau-level mixing can thus tip the balance oppositely at $\nu=\frac{1}{4}$ and $\nu=\frac{3}{4}$. According to the phase diagram in Fig.~\ref{fig.phase_quart}, such interactions are most likely to favor the anti-Pfaffian phase, which is the closest to the 0LL Coulomb point. This result is compatible with a satellite plateau that has been observed at the filling factor $\nu=\frac{19}{25}$, consistent with anti-Pfaffian topological order.~\cite{Yutushui_daughters_2024} 

The method we developed readily applies to any single-component quantum Hall system. In particular, it enables Monte Carlo studies of arbitrary Landau levels in monolayer and multilayer graphene with physically motivated perturbations. It would be valuable to generalize this method to multicomponent systems and study the competition between polarized and singlet quantum Hall states.

Another promising application of this technique is the construction of model interactions that target specific topological orders. Given a trial state, one can compute pseudopotential energies and use them to construct favorable interactions. This approach could be especially valuable for bosons, where Fock-space representations of a trial state cannot be easily obtained via Monte Carlo.

Finally, we expect that one could generalize our approach to many-body interactions. Specifically, three-body pseudopotentials are commonly used in exact diagonalization studies of Landau level mixing effects. Computing their expectation values with Monte Carlo would enable studying the effects of Landau level mixing on larger systems. We note, however, that Landau level mixing does not always break particle-hole symmetry, e.g., in the zeroth Landau level of graphene with electrons and holes excited symmetrically. In such cases, the three-body interactions vanish, and our method can be readily used to assess the effects of Landau level mixing on energies.

\textit{Note added---}The results of Ref.~\onlinecite{Huang_3_4_2024}, which studied the $\nu=3/4$ filled states in bilayer graphene, are consistent with our finding. A recent study Ref.~\onlinecite{Milovanovic_pairing_2024} relates pairing between composite fermions to interactions between their dipole moments.

\begin{acknowledgments}
It is a pleasure to thank Ajit Balram for illuminating conversations and sharing some of his unpublished results. This work was supported by the Israel Science Foundation (ISF) under grant 2572/21 and by the
Minerva Foundation with funding from the Federal German
Ministry for Education and Research.
\end{acknowledgments}
 \section*{DATA AVAILABILITY}
 Data used in the paper is publicly available in Zenodo at \href{https://zenodo.org/records/14030395}{10.5281/zenodo.14030395}.

\label{sec.discussion}

\appendix
\section{Higher Landau levels}\label{app.higher}
The energy of Coulomb interactions between electrons in the $n$th Landau level can be computed from the lowest Landau level wavefunction with modified interactions $V^\text{$n$LL}_k$ by Eq.~\eqref{eq.energy}. We now find the harmonics  $V^\text{$n$LL}_k$ for arbitrary interactions parametrized by $V_k$ in Eq.~\eqref{eq.expnasion2}.

On a sphere with monopole strength $q'=l-n$, the number of states in the $n$th Landau level is $2l+1$. Consequently, any two-body interactions in this Landau level are parametrized by $2l+1$ Haldane pseudopotentials $H^\text{$n$LL}_L$ (cf. the main text, where $2q+1$ pseudopotentials $H_L$ parameterize the lowest Landau level interactions). Given the harmonics $V_k$, these pseudopotentials can be  computed~\cite{wooten_haldane_2013} as
\begin{align}\label{eq.app_V2H_1}
    H^\text{$n$LL}_L = \sum_{k=0}^{2l}M_{L,k}(l,q'=l-n) V_k.
\end{align}
The $2l+1$-dimensional transition matrix $M_{L,k}$ is defined via
\begin{align}\label{eq.app_M_mat}
    M_{L,k}(l,q')\equiv [N_{k,L}(l,q')]^{-1} =(-1)^{-L}(2l+1)^2\\\times
\left(
\begin{array}{ccc}
 l & k & l \\
 -q' & 0 & q' \\
\end{array}
\right)^2
    \left\{
\begin{array}{ccc}
 2l-L & l & l \\
 k & l & l \\
\end{array}
\right\}~,
\end{align}
where $N(l,q')$ generalizes the matrix $N(q=l)$ in Eq.~\eqref{eqn.nkl} of the main text and is given by
\begin{align}
    N_{k,L}(l,q) =(-1)^L\frac{(2k+1)}{(2l+1)^2}(2(2l-L)+1\nonumber\\\times
\left(
\begin{array}{ccc}
 l & k & l \\
 -q & 0 & q \\
\end{array}
\right)^{-2}
    \left\{
\begin{array}{ccc}
 l & l & k\\
 l & l & 2l-L  \\
\end{array}
\right\}.
\label{eqn.nqkl}
\end{align}

We determine $V^\text{$n$LL}_k$ that reproduces $n$th Landau level interactions in the lowest Landau level. To match the number of states in $n$LL, the 0LL occurs on a sphere with $q=l=q'+n$, and thus 
\begin{align}\label{eq.app_V2H_2}
    V^\text{$n$LL}_k = \sum_{L=0}^{2l}N_{k,L}(l,q=l) H^\text{$n$LL}_L.
\end{align}
We insert Eqs.~\eqref{eq.app_V2H_1} into \eqref{eq.app_V2H_2}, to obtain the  harmonics 
\begin{align}
    V^\text{$n$LL}_{k'} = \sum_{L,k} N_{k',L}(l,q) M_{L,k}(l,q') V_k.
\end{align}
 Using the orthogonality relation of the 6$j$-symbols, we find that $\sum_{L=0}^{2l} N_{k,L}(l,q) M_{L,k'}(l,q')\propto \delta_{k,k'}$ and obtain the real-space representation of $V(\theta)$ projected into the $n$th Landau level 
\begin{align}\label{eq.app_Vk$n$LL}
    V^\text{$n$LL}_k=V_k
\left(
\begin{array}{ccc}
 l & k & l \\
 -q' & 0 & q' \\
\end{array}
\right)^{2}
    \left(
\begin{array}{ccc}
 l & k & l \\
 -l & 0 & l \\
\end{array}
\right)^{-2}.
\end{align}
For the two lowest Landau levels, this equation reduces to
\begin{align}
    V^\text{0LL}_k=V_k,\qquad
    V^\text{1LL}_k=V_k \left(\frac{2l-k(k-1)}{2l}\right)^2.
\end{align}
The harmonics $V^\text{$n$LL}_k$ grow as $k^{4n} V_k$, and a refinement scheme in Section~\ref{sec.refine} becomes necessary for numerical stability. 
 
\begin{figure}
    \centering
    \includegraphics[width=0.49\linewidth]{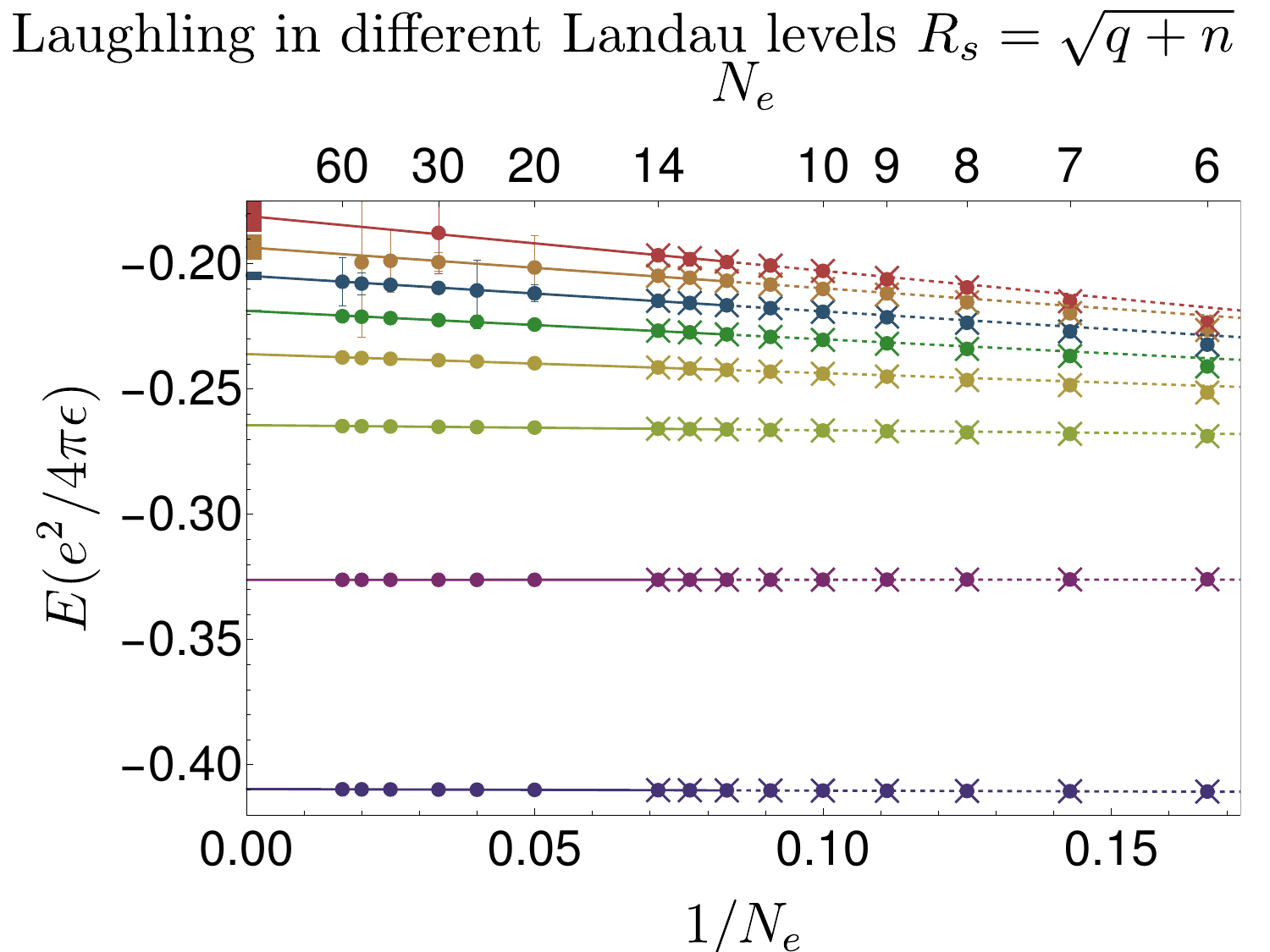}
    \includegraphics[width=0.49\linewidth]{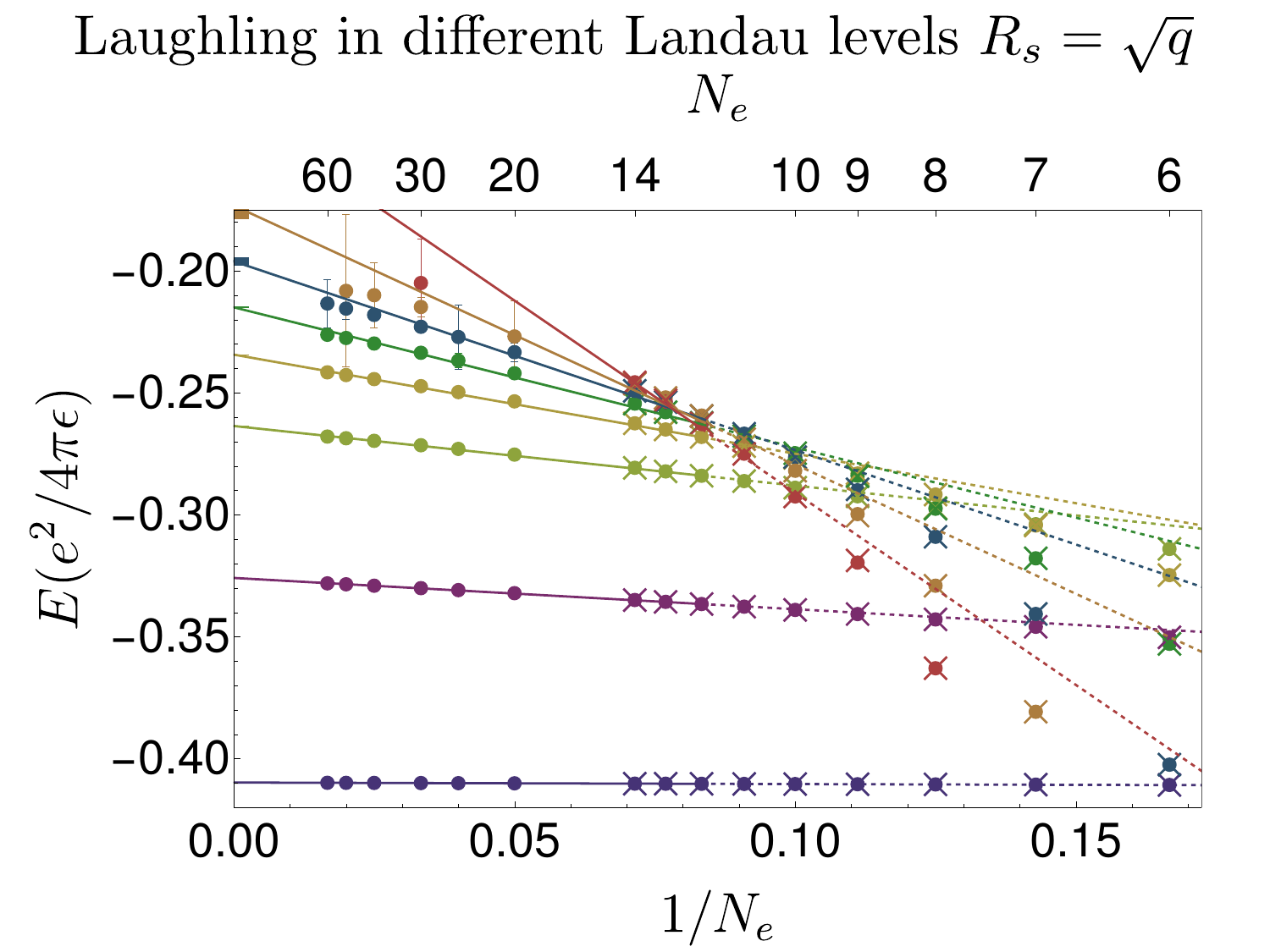}
    \caption{Comparison of different conventions for the radius of a sphere $R_s = \sqrt{q+n}$ and $R_s'=\sqrt{q}$. In the latter case, finite-size effects are considerably stronger in higher Landau levels and allow linear extrapolation only starting from much larger systems.}
    \label{fig.Laughling_Rs}
\end{figure}

\section{Background and finite-size corrections}\label{app.background}
To extrapolate to the thermodynamic limit, we subtract the background contribution from the evenly distributed neutralizing charge $|e|N_e$. Different conventions for this subtraction lead to the same energy in the thermodynamic limit but differ at finite $N_e$. A natural choice is to assume Coulomb interactions among these charges, leading to the background energy $E_\text{bg} = \frac{N^2_e}{2}V^\text{$n$LL}_0 = \frac{N^2_e}{2 R_s}$, independent of Landau level $n$  [cf.~Eq.~\eqref{eq.Vk$n$LL}]. However, the energy extrapolates better to the thermodynamic limit~\cite{Sharma_CF_pairing_quart_2024} if we take the same interaction potential $V(\theta)$ for the background charge and electrons and subtract  $ E_\text{bg} =\frac{ N_e^2}{2} V_0$ from Eq.~\eqref{eq.energy}. Notice that we take the original value of $V_0$ and not the refined $\tilde V_0$. The latter are merely used to mitigate statistical error.

We rescale the energy by the factor $\alpha = \sqrt{\frac{ 2l\nu}{ N_e}}$ to take into account the shift of different states.~\cite{Jain_composite_2007} Here, $2l$ is the number of fluxes for the lowest Landau level wavefunction. Finally, we take $R_s =\sqrt{l}$ independent of the Landau level index $n$ to keep the average inter-particle spacing fixed. An alternative convention of taking $R'_s = \sqrt{q}=\sqrt{l-n}$ makes finite-size effects considerably stronger see Fig.~\ref{fig.Laughling_Rs}. The final energy we plot in the main text is 
\begin{align}\label{eq.energy_fin}
    E_\text{int} =\frac{\alpha N_e^2}{2\sqrt{l}} \left(\sum_{k=0}^{2l} \frac{\tilde{V}_kG_k}{2k+1} - V_0\right).
\end{align}

\begin{figure}[t!]
    \centering
    \includegraphics[width=1\linewidth]{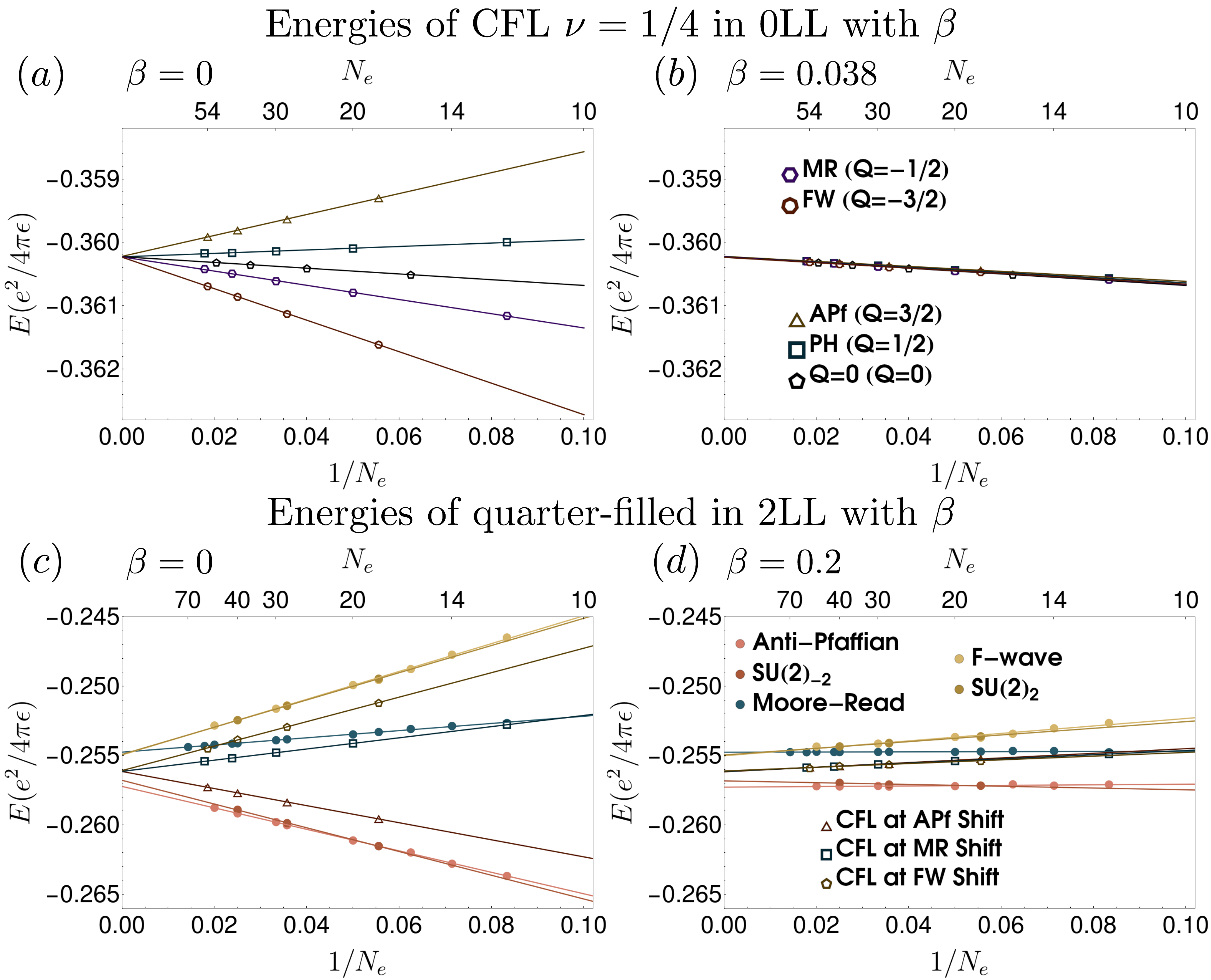}
    \caption{ The thermodynamic fit of the energies with conventional normalization $\beta=0$ and tuned parameter $\beta\neq 0$. Composite Fermi liquids at quarter filling in 0LL with (a) $\beta=0$ and (b) $\beta=0.038$. Quarter-filled states in 2LL with (c) $\beta=0$ and (d) $\beta=0.2$. The thermodynamic energies are independent of $r$ for all states; however, the finite-size effects are greatly mitigated by a single parameter of the fitting procedure $\beta$.}
    \label{fig.rescaled_1}
\end{figure}

The finite system size dependence can be further reduced by introducing an {\it interaction} and filling-factor-dependent phenomenological constant $\beta$ that does not change the thermodynamic energies. We rescale the energy Eq.~\eqref{eq.energy_fin} by 
\begin{align}
    E_\text{int} \to E_\text{int} \left(1  - \frac{\beta Q}{N_e}\right)
\end{align}
and tune the shift-independent number $\beta$ such that the linear fits in $N^{-1}_e$ of different CFLs collapse; see  Fig.~\ref{fig.rescaled_1}. The coefficient $\beta$ can be inferred from the Monte Carlo simulation of known states, e.g., CFL. The improved normalization can facilitate easier comparison of smaller systems studied via exact diagonalization. In Fig.~\ref{fig.rescaled_1}, we show that the resulting linear fits for all paired states exhibit much weaker finite-size effects (their slopes become similar).


\begin{table}[b]
  \centering
  \renewcommand{\arraystretch}{1.2} 
  \caption{Squared overlaps $\langle\Psi_\text{P-1}|\Psi_\text{P-2}\rangle$ for different pairing channels projected with different schemes measured in percentages. The large overlaps for moderate systems suggest that the states represent the same phase, and either scheme can be used. For the f-wave state with $N_e=20$, the discrepancy is the largest.} 
  \begin{tabular}{|c| c c c c|}
    \hline  ($N_e,N_c$) & FW($\ell=3$) & MR($\ell=1$) & PH($\ell=-1$) & aPf($\ell=-3$)
    \\ \hline\hline
    (10,10) & $0.731(7\pm5) $ & $0.9975(5\pm1)$ & $0.90(0\pm1)$ & $0.92(8\pm1)$ \\
    (12,10) & $0.74(0\pm2)$ & $0.9920(0\pm4)$ & $0.935(7\pm4)$ & $0.91(5\pm2)$ \\
    (20,14) & $0.6(3\pm 3)$ & $0.988(6\pm 4)$ & $0.868(3\pm 4)$ & $0.8(0\pm 2)$ \\ \hline
  \end{tabular}
  \label{tab.app_overlaps}
\end{table}

\section{Harmonics for individual Haldane pseudopotentials}\label{app.higher_pot}
\begin{figure}
    \centering
    \includegraphics[width=\linewidth]{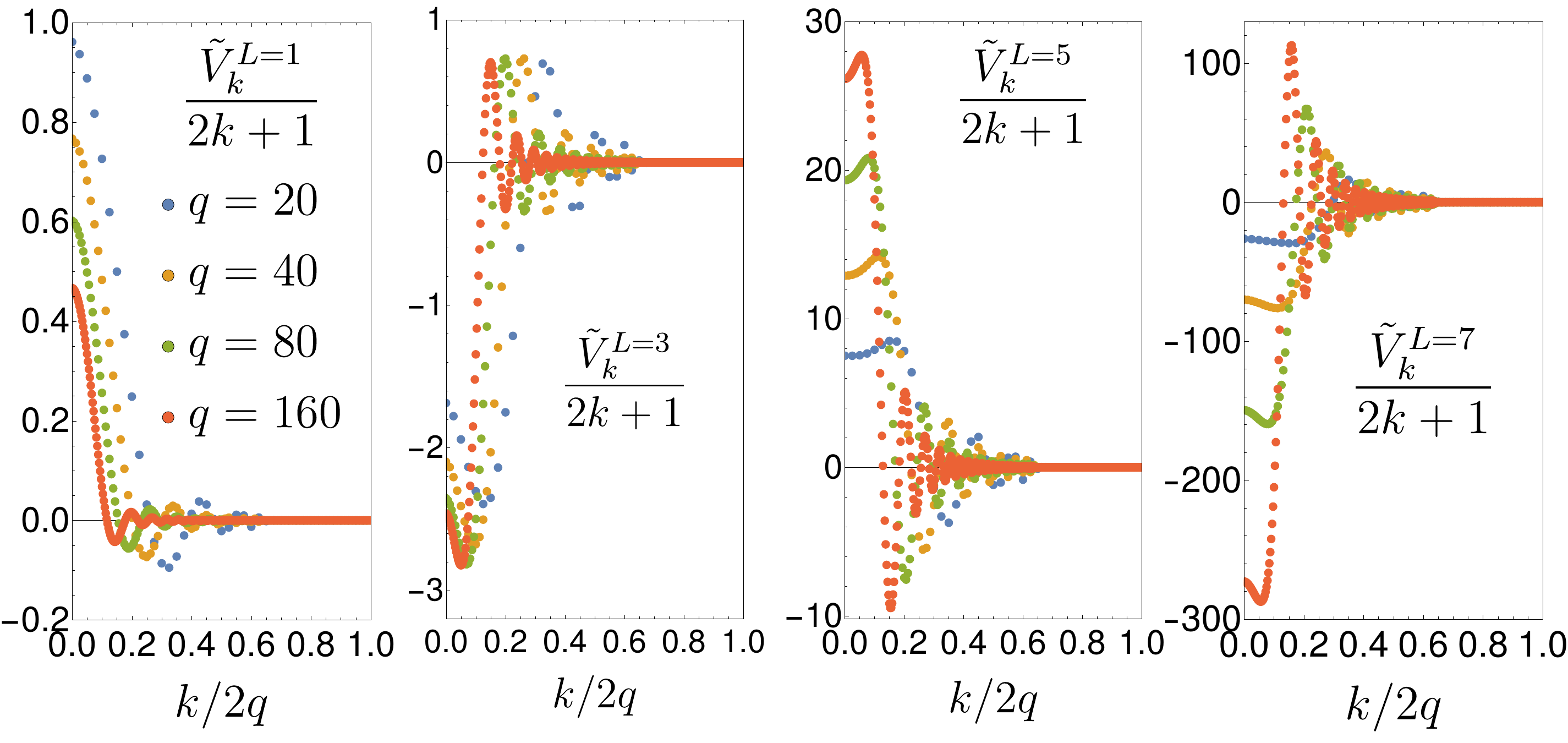}
    \caption{The refined harmonics for individual pseudopotentials $H_{L}=H^{0LL}_{L_0}\delta_{L,L_0}$ for $L_0=1-7$ and different systems sizes $q$. The small-$k$ harmonics grow rapidly with $L_0$, causing larger noise in the resulting energies. For large $L$, the small-$k$ harmonics depend strongly on the system size, rendering pure convergence for the largest systems.}
    \label{fig.Haldane_higher} 
\end{figure}

In the main text, we identified the primary source of the noise to be the rapid growth of $V_k$ with large $k$. To this end, we introduced the refining scheme that suppressed large-$k$ harmonics $V_k$ at the expense of making small-$k$ harmonics larger. For individual Haldane pseudopotentials  $H_{L}=H^{0LL}_{L_0}\delta_{L,L_0}$ with large $L$, the small-$k$ harmonics rapidly grow with the system size and $L$. In Fig.~\ref{fig.Haldane_higher}, we show the refined $\tilde{V}_k$ for individual Haldane pseudopotentials. Therefore, the method introduced in the paper works best for the lowest Haldane pseudopotentials, which are also the most relevant ones.

\section{Landau-level projection}\label{app.projection}


\subsection{Paired composite fermion states}
In the main text, we study paired states at $\nu=\frac{1}{4}$ with pairing channels corresponding to f-wave $(\ell=3)$, Moore-Read ($\ell=1$), and anti-Pfaffian ($\ell=-3$). The wavefunctions we use are
\begin{align}\label{eq.sccf}
    \Psi_\text{SCCF} = P_\text{LLL}\left\{ \text{Pf}\left(\frac{z_{ij}^{Q-1/2}}{[z^*_{ij}]^{Q+1/2}}\right) \prod_{i<j}z_{ij}^4 \right\},
\end{align}
where $\ell=-2Q$, $z_{ij}=z_i-z_j$, and Pf denotes Pfaffian, which can be efficiently evaluated.~\cite{pfapack_Wimmer_2012} In spherical geometry~\cite{Haldane_fqh_1983}, we replace $z_{ij}\to\omega_{ij}=u_iv_j-u_jv_i$ with $(u_i,v_i)=(\cos\frac{\theta}{2},\sin\frac{\theta}{2})e^{i\frac{\phi}{2}}$. The pairing function can be expanded~\cite{Yutushui_Large_scale_2020} in terms of monopole harmonics~\cite{Wu_dirac_1976,Wu_properties_1977} as
\begin{align}
    \frac{\omega_{ij}^{Q-1/2}}{[\omega^*_{ij}]^{Q+1/2}} &= 2\omega_{ij}^{2Q}\sum_{n=0}^{\infty} P_n^{(2Q,0)}(1-2|\omega_{ij}|^2)
    \\&=
    \sum_{l=|Q|}^{\infty}\frac{8\pi}{2l+1} \sum_{m=-l}^{l}(-1)^{Q+m}Y^{Q}_{l,m}(r_i)\bar{Y}^{Q}_{l,m}(r_j), \notag
\end{align}
where $\bar{Y}^{Q}_{l,m} \equiv (Y^{-Q}_{l,m})^*$. Absorbing the Jastrow factor $\prod_{i<j}\omega_{ij}^{4}$ into the Pfaffian, the paired wavefunction  can be written as
\begin{align}
     \Psi_{\text{SCCF}} =  P_\text{LLL}\text{Pf}
     \left(
     \sum_{l,m}^{\infty}\eta_l \left\{ {Y}^{Q}_{l,m}(r_i) 
     J_i^2 \right\}\left\{ \bar{{ Y}}^{Q}_{l,m}(r_j)  J_j^2
     \right\}
     \right),
\end{align} 
where  $\eta_l=\frac{8\pi}{2l+1}$, and $J_i=\prod_{a\neq i}\omega_{ai}$ attaches one flux quantum to the $i$th particle. The lowest-Landau-level projection can be performed efficiently by approximating $\Psi_{\text{SCCF}}$ with  
\begin{align}
     \Psi_\text{P-1} =  \text{Pf}
     \left(
     \sum_{l,m}^{N_c+|Q|}\eta_l \;
     {}^2{\cal Y}^{Q}_{l,m}(r_i)\;
     {}^2\bar{{\cal Y}}^{Q}_{l,m}(r_j)
     \right),
\end{align} 
where the projection acts separately on the composite-fermion orbitals, i.e., $^p{\cal Y}^{Q}_{l,m}(r_i)\equiv P_\text{LLL}[Y^{Q}_{l,m}(r_i) J^p_i]$. The calculation of ${}^2{\cal Y}^{Q}_{l,m}(r_i)$ can be done following Refs.~\onlinecite{Davenport_projection_2012,Fulsebakke_projection_2016,Mukherjee_incompressible_2015}. A significant simplification is to further approximate $ {}^2{\cal Y}^{Q}_{l,m}(r_i)\to{}^1{\cal Y}^{Q}_{l,m}(r_i) J_i$. Then, the wavefunction can be projected much faster as
\begin{align}\label{eq.app_sch2}
     \Psi_\text{P-2} =  \text{Pf}
     \left(
     \sum_{l,m}^{N_c+|Q|}\eta_l 
    \;{}^1{\cal Y}^{Q}_{l,m}(r_i)\; {}^1\bar{{\cal Y}}^{Q}_{l,m}(r_j)
     \right)\Phi_1^2.
\end{align} 
\begin{figure}[b]
    \centering
    \includegraphics[width=1\linewidth]{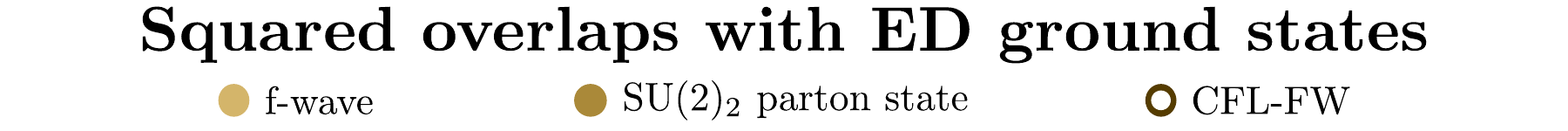}\\
    \includegraphics[width=1\linewidth]{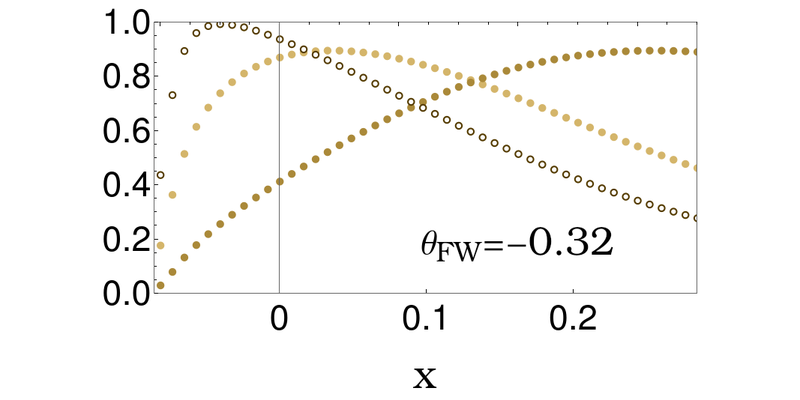} 
    \caption{The overlaps of the ground state $|GS(x)\rangle$ of Eq.~\eqref{eqn:hx} with the trial states for CFL, f-wave state, and $SU(2)_2$ parton state Eq.~\eqref{eq.app_SU22} at $\nu=\frac{1}{4}$ for $N_e=10$.}
    \label{fig.overlaps_fw}
\end{figure}

To justify this approximation, we compute overlaps of $\Psi_\text{P-1}$ and $\Psi_\text{P-2}$ for different pairing channels, see Tab.~\ref{tab.app_overlaps}. The largest discrepancy arises for the f-wave pairing channel. However, even for moderately large systems with $N_e=20$ electrons at {\it quarter filling}, the smallest squared overlap is $0.6$. The 0LL Coulomb energies of $\Psi_\text{P-1}$ and $\Psi_\text{P-2}$ are $-0.358(1\pm 2)$ and $-0.35950(5\pm6)$, respectively. The second projection scheme yields somewhat lower energies due to $\Phi_1^2$ that results in vanishing $E_{L=1}=0$ (cf. Fig.~\ref{fig.Haldane_quart}) and is numerically more efficient. We, therefore, used $\Psi_\text{P-2}$ for all data in the main text.

\begin{table}[t!]
  \centering
  \renewcommand{\arraystretch}{1.2} 
  \caption{Squared overlaps of different wavefunctions for the $f$-wave state computed with exact diagonalization. The FW, FWS and $SU(2)_{2}$ denotes $\Psi^{\nu=1/2}_\text{FW}$, $\Psi^{\nu=1/2}_\text{FWS}$ and $\Psi^{\nu=1/2}_{SU(2)_{2}}$, respectively.}
  \begin{tabular}{|c c| c c c|}
    \hline  $N_{e}$ & dim & $\langle SU(2)_{2}|\text{FW}\rangle$ & $\langle SU(2)_{2}|\text{FWS}\rangle$ & $\langle \text{FW}|\text{FWS}\rangle$
    \\ \hline\hline
    8 & 33 &  0.977011 & 0.991829 & 0.996194 \\
    10 & 338 & 0.902073 & 0.989419  & 0.948236  \\
    12 & 3788 & 0.901462 & 0.986931 & 0.945853 \\
    14 & 44916 & 0.884793 & 0.984970 & 0.934026 \\
    \hline
  \end{tabular}
  \label{tab.app_overlaps3}
\end{table}


\subsection{Parton wavefunction}
The alternative trial state for the anti-Pfaffian phase,~\cite{Balram_parton_2018}  Eq.~\eqref{eqn.parton} for $SU(2)_{-2}$,  can be efficiently projected at half- and quarter filling by approximating 
\begin{align}\label{eq.asu22}
    \Psi^{\nu=1/2p}_{SU(2)_{-2}} = P_\text{LLL} \Phi_{-2}^2 \Phi^{2p+1}_1\approx 
    \left[P_\text{LLL} \Phi_{2}\Phi_1^{2}\right]^2\Phi_1^{2p-3}~.
\end{align}
The factor $\Phi_1^{-1}$ for half filling ($p=1$) does not cause numerical instability. However, the parton ansatz for f-wave pairing, Eq.~\eqref{eqn.parton} for $SU(2)_{2}$, can be efficiently projected only at quarter filling
\begin{align}\label{eq.app_SU22}
    \Psi^{\nu=1/4}_{SU(2)_{2}} = P_\text{LLL} \Phi_{2}^2 \Phi_1^{3}\approx 
    \left[P_\text{LLL} \Phi_{2}\Phi_1^{2}\right]^2\Phi_1^{-1}~.
\end{align}
For the half-filled state, a similar approximation would lead to the factor $\Phi_1^{-3}$, which leads to instability. Nevertheless, all four states can be projected by computing the basis overlaps to obtain Fock-space wavefunctions for small systems. 

\begin{figure}[t]
    \centering
    \includegraphics[width=0.8
    \linewidth]{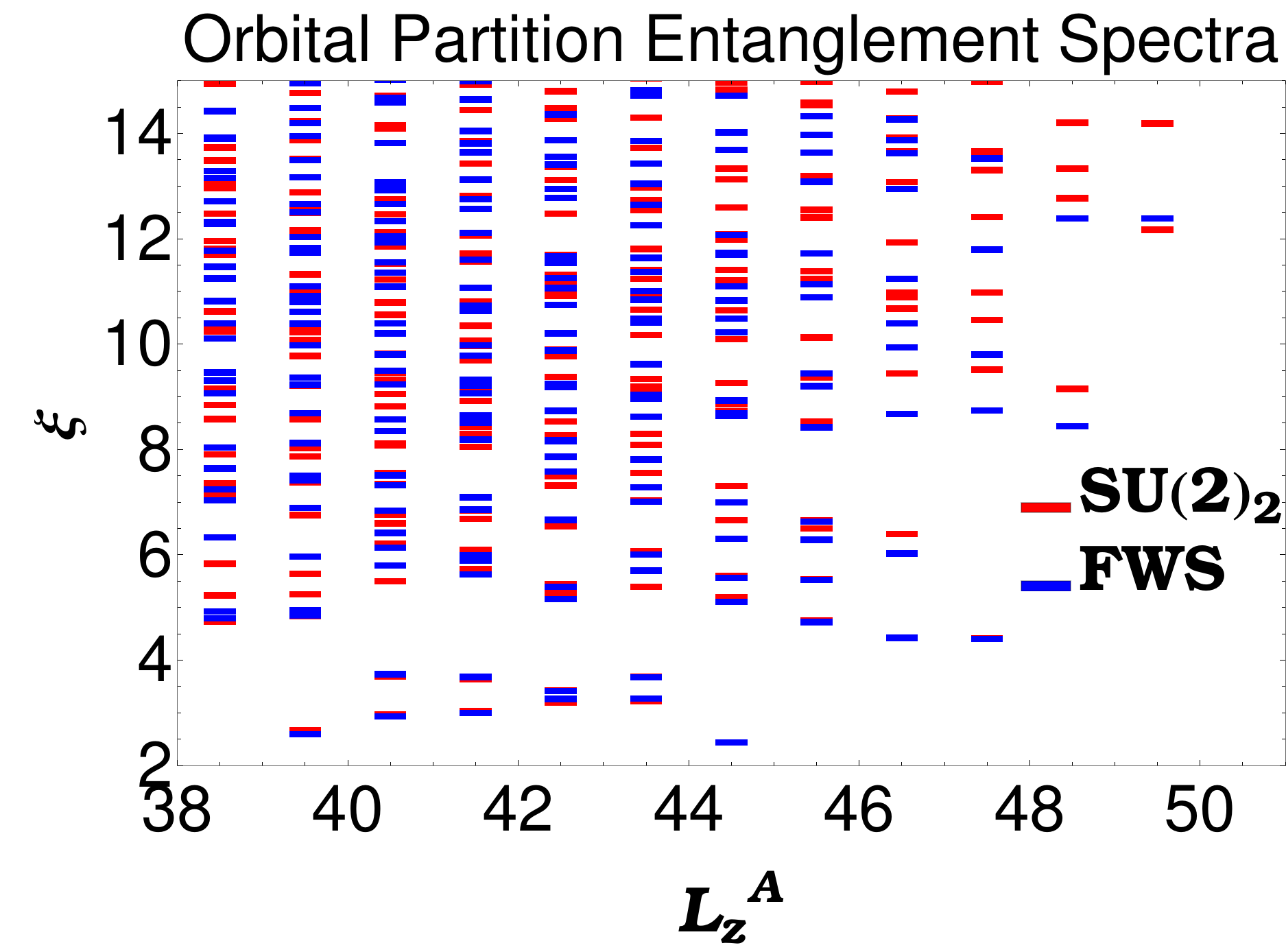}\\
    \caption{Orbital partition entanglement spectra~\cite{li_entanglement_2008} of the two wavefunctions for the $f$-wave state for $N_e=14$ with equal partitioning. The low-lying branches match each other well, demonstrating that the wavefunctions are in the same phase. 
    }
    \label{fig.ES}
\end{figure}

In Ref.~\onlinecite{Balram_parton_2018}, the wavefunction $\Psi^{\nu=1/2}_{SU(2)_{-2}}$ is shown to be a good approximation of the particle-hole conjugate of Moore-Read and represents the anti-Pfaffian phase. Here we provide evidence that the parton wavefunction $\Psi^{\nu=1/2}_{SU(2)_{2}}$ is, in fact, a paired state with the $f$-wave pairing channel. In Tab.~\ref{tab.app_overlaps3}, we show the overlaps of $\Psi^{\nu=1/2}_{SU(2)_{2}}$ with the $f$-wave wavefunction in Eq.~\eqref{eqn.apr}, i.e., $\Psi^{\nu=1/2}_\text{FW}\equiv \Psi_{\ell=3}(p=1)$ and its single-particle lowest Landau level projection version [using Eq.~\eqref{eq.app_sch2}] at half-filling, i.e., $\Psi^{\nu=1/2}_\text{FWS} = \Psi_\text{P-2}/\Phi_1^2$. The exceptionally large overlap between the three states for moderate systems suggests that they represent the same phase; see also Fig.~\ref{fig.overlaps_fw}. To further show that these states represent the same state, we compute the entanglement spectra for $\Psi^{\nu=1/2}_\text{FWS}$ and $\Psi^{\nu=1/2}_{SU(2)_{2}}$ and plot it in Fig.~\ref{fig.ES}. Therefore, we conclude that $SU(2)_2$ and $f$-wave wavefunction represent the same phase, and both are valid trial states for the $f$-wave paired phase of composite fermions. Nevertheless, the microscopic details of the wavefunction are slightly different. Consequently, their energies are different, and we include both in our study.

\begin{table}
  \centering
  \renewcommand{\arraystretch}{1.2} 
  \caption{Squared overlaps of paired states and CFLs at half filling.}
  \begin{tabular}{|c| c c c c|}
    \hline  $N_{F}$ & FW($\ell=3$) & MR($\ell=1$)& PH($\ell=-1$) & aPf($\ell=-3$)
    \\ \hline\hline
    1 & $0.5645(9\pm 6)$ & $0.864(1\pm 2)$ & $0.972(5\pm 1)$ & $0.451(9\pm 1)$ \\
    2 & $0.202(2\pm 2)$ & $0.492(7\pm 3)$ & $0.846(7\pm 1)$ & $0.154(4\pm 5)$ \\
    3 & $0.050(4\pm 2)$ & $0.183(5\pm 3)$ & $0.579(4\pm 3)$ & $0.037(6\pm 2)$ \\
    4 & $0.008(9\pm 2)$ & $0.048(0\pm 2)$ & $0.293(2\pm 8)$ & $0.005(8\pm 1)$ \\
    \hline
  \end{tabular}
  \label{tab.app_overlaps2}
\end{table}

\begin{table}
  \centering
  \renewcommand{\arraystretch}{1.2} 
  \caption{Squared overlaps of paired states and CFLs at quarter filling.}
  \begin{tabular}{|c| c c c c|}
    \hline  $N_{F}$ & FW($\ell=3$) & MR($\ell=1$) & PH($\ell=-1$) & aPf($\ell=-3$)
    \\ \hline\hline
    1 & $0.7530(6\pm 5)$ & $0.9435(8\pm 2)$ & $0.99176(1\pm2)$ & $0.632(1\pm 1)$ \\
    2 & $0.377(4\pm 3)$ & $0.686(6\pm 2)$ & $0.9427(4\pm 4)$ & $0.287(8\pm 3)$ \\
    3 & $0.131(8\pm 4)$ & $0.345(9\pm 2)$ & $0.792(8\pm 2)$ & $0.094(7\pm 6)$ \\
    4 & $0.034(0\pm 3)$ & $0.125(9\pm 3)$ & $0.540(4\pm 4)$ & $0.024(1\pm 3)$ \\
    \hline
  \end{tabular}
  \label{tab.app_overlaps1}
\end{table}

\section{Overlap of paired states with CFL}\label{app.overlaps}
We compute the overlaps of the paired state at half and quarter filling, Eq.~\eqref{eq.app_sch2}, with the CFL at corresponding shifts in Tabs.~\ref{tab.app_overlaps2} and~\ref{tab.app_overlaps1}.

\bibliography{ref}

\begin{thebibliography}{91}
\bibitem [{foo()}]{footnote}%
  \BibitemOpen
  \bibfield  {title} {\enquote {\bibinfo {title} {We do not include $k=0$ contribution to the optimization function $\mathcal{F}$ since {$G_{k=0}=\frac{N_e-1}{N_e}$} is known analytically and does not cause noise in computed energy. however, including $k=0$ in the function does not change the result significantly.}}\ }\href@noop {} {\ }\BibitemShut {NoStop}%
\end{thebibliography}

\end{document}